\documentclass[a4paper,10pt,fleqn,usenames,dvipsnames]{article}

\usepackage{lmodern}
\usepackage[T1]{fontenc}
\usepackage{amsmath}
\usepackage{amssymb}

\usepackage{url}
\usepackage{longtable}
\usepackage{multirow}
\usepackage{caption}
\usepackage[labelformat=simple]{subcaption}

\usepackage{etoolbox} % Add this if not already included

\AtBeginEnvironment{appendix}{%

}

\usepackage[nodayofweek]{datetime}
\usepackage{enumerate}
\usepackage{xspace}
\usepackage{xcolor}
\usepackage{graphicx}
\usepackage{bookmark}
\usepackage{pgffor}
\usepackage{comment}

\usepackage{float}
\usepackage[capitalise]{cleveref}

\usepackage[backend=biber,giveninits=true,doi=false,sorting=none,style=numeric-comp,maxnames=10]{biblatex}
\DeclareFieldFormat[article]{citetitle}{\textit{#1}\isdot}
\DeclareFieldFormat[article]{title}{\textit{#1}\isdot}
\DeclareFieldFormat[unpublished]{citetitle}{\textit{#1}\isdot}
\DeclareFieldFormat[unpublished]{title}{\textit{#1}\isdot}
\DeclareFieldFormat[inproceedings]{citetitle}{\textit{#1}\isdot}
\DeclareFieldFormat[inproceedings]{title}{\textit{#1}\isdot}

\DeclareNameAlias{author}{last-first}

\newcommand{\DL}{\ensuremath{\mathcal{D}_L}\xspace}

\newcommand{\OMLP}{MLP Head\xspace}

\newcommand{\psim}{\ensuremath{p_{\text{sim}}}\xspace}

\newcommand{\pdata}{\ensuremath{p_{\text{data}}}\xspace}
\newcommand{\CR}{\text{CR}}
\newcommand{\omegap}{\ensuremath{\omega^\prime}}

\newcommand{\Neff}{\ensuremath{N_\mathrm{eff}}\xspace}
\newcommand{\Nsim}{\ensuremath{N_\mathrm{sim}}\xspace}

\usepackage[nonatbib,preprint]{neurips}

\title{Mind the Gap:\\ Navigating Inference with Optimal Transport Maps}
\author{%
  Malte Algren${}^*$ (\texttt{malte.algren@unige.ch}), Tobias Golling (\texttt{tobias.golling@unige.ch}) \\
  University of Geneva \\
  \And
  Francesco Armando Di Bello${}^*$ (\texttt{francesco.armando.di.bello@unipi.it}) \\
  INFN and University of Pisa \\
  \And
  Christopher Pollard${}^*$ (\texttt{christopher.pollard@warwick.ac.uk}) \\
  University of Warwick \\
}

\begin{document}
    \maketitle

    \begin{abstract}
      Machine learning (ML) techniques have recently enabled enormous gains in
      sensitivity to new phenomena across the sciences.
      In particle physics, much of this progress has relied on excellent
      simulations of a wide range of physical processes.
      However, due to the sophistication of modern machine learning algorithms
      and their reliance on high-quality training samples, discrepancies between
      simulation and experimental data can significantly limit their
      effectiveness.
      In this work, we present a solution to this ``misspecification'' problem:
      a model calibration approach based on optimal transport, which we apply to
      high-dimensional simulations for the first time.
      We demonstrate the performance of our approach through jet tagging, using
      a simulated dataset inspired by the Compact Muon Solenoid experiment at
      the Large Hadron Collider. 
      A 128-dimensional internal jet representation from a powerful
      general-purpose classifier is studied; after calibrating this internal
      ``latent'' representation, we find that a wide variety of quantities
      derived from it for downstream tasks are also properly calibrated: using
      this calibrated high-dimensional representation, powerful new applications
      of jet flavor information can be utilized in LHC analyses.
      This is a key step toward allowing the unbiased use of ``foundation
      models'' in particle physics.
      More broadly, this calibration framework has broad applications for
      correcting high-dimensional simulations across the sciences.
    \end{abstract}

    {
      \renewcommand{\thefootnote}{\fnsymbol{footnote}}
      \footnotetext[1]{Corresponding authors}
    }

    \clearpage

    \section{Introduction}
    
The use of inference techniques built on complex and precise statistical models,
often enabled by sophisticated simulations, continues to become more widespread
across the sciences.
This has led to a recent boom of so-called ``simulation-based inference`` (SBI),
in which neural methods are used to perform inference over information-rich 
observations~\cite{doi:10.1073/pnas.1912789117,zammitmangion2024neuralmethodsamortizedinference,Brehmer:2021ivt,wang2024comprehensiveguidesimulationbasedinference}. 
When the family of statistical models used for inference does not cover the
generation process of observed data, a misspecification gap emerges, and
constraints on model parameters will no longer be robust.
Indeed, there is a growing body of work quantifying the impact of
misspecification in a variety of
environments~\cite{cannon2022investigatingimpactmodelmisspecification} and in
particular in the context of
SBI~\cite{montel2025testsmodelmisspecificationsimulationbased,schmitt2022detectingmodelmisspecificationamortized,huang2023learningrobuststatisticssimulationbased}.
Given the power of neural methods to gracefully handle high-dimensional and
informative observations, it is critical to develop techniques that mitigate
misspecification, enabling inference that is both effective and
unbiased~\cite{CranmerEtAl2015}.

In the presence of control data, optimal transport (OT)
theory~\cite{monge1781memoire,villani2008optimal} provides a framework for
addressing the misspecification problem through model calibration.
Pollard and Windischhofer~\cite{Pollard_2022} argued that OT is an appropriate
solution to model misspecification, and the ATLAS
collaboration~\cite{ATLAS-PAPER-2025} successfully applied this strategy to
calibrate three-dimensional simulations with particle collider data from the
Large Hadron Collider~\cite{Evans:2008zzb}.
A similar approach was subsequently applied to calibrate models for
cardiovascular
health~\cite{wehenkel2025addressingmisspecificationsimulationbasedinference}.

In this work, we show for the first time both in simplistic and realistic
examples that the OT-based calibration procedure scales well to high dimensions
for an important class of hierarchical models.
We find that density ratio-based corrections to samples drawn from a forward
model are unsuitable, even in simple cases, for a large body of
high-dimensional problems.
We additionally show that, in a familiar scenario from particle physics, the
densities between a baseline forward model and simulated data are not
sufficiently overlapping for density-ratio corrections to be feasible.
Moreover, we derive OT solutions to address the misspecification gap in the same
particle-physics example and conclude that the model misspecification is
adequately corrected in an inference-relevant way.
This result lays the groundwork for properly-calibrated foundation models in
scenarios with suitable calibration data, which enables their use for
high-performance, but unbiased, statistical inference.

    \section{Calibration of hierarchical models}
    In hierarchical models, the likelihood of observing some datum $x$, given a
parameter of interest $\theta$, is determined by marginalizing over any
stochastic latent variable $\omega$:
$p(x|\theta) = \int d\omega\, p(x|\omega, \theta)\, p(\omega|\theta)$.
An important subset of hierarchical models comprises those which
factorize, such that $x$ only depends on $\theta$ through $\omega$, i.e.
$p(x|\omega, \theta) = p(x|\omega)$.
In such situations, it may be possible to calculate $p(x|\omega)$ separately, or
to constrain it via auxiliary experiments.
Indeed, this is often the motivation for constructing the likelihood
$p(x|\theta)$ in terms of a stochastic latent $\omega$: when
$p(x|\theta) \approx \int d\omega\, p(x|\omega)\, p(\omega|\theta)$, then both
$p(x|\omega)$ and $p(\omega|\theta)$ can be independently derived or verified.

Inference on approximately factorizable likelihoods is especially powerful.
Scientific instruments are often designed to measure a quantity $x$ regardless
of the details of the particular experiment, except for the stochastic latent
variable $\omega$.
For example, when measuring the mass of an object, an ideal scale is one that
provides an accurate and precise estimate of the object's weight,
$p(x = \text{estimated mass}\, |\, \omega = \text{true mass})$,
independently of whether the object is, say, a penguin or a banana.
A scale with this factorization property can be calibrated using a set of test
objects whose weights are well-known; in doing so, the experimenter constructs a
good approximation of the scale's density over outputs for a known input.
An object's mass can then be used to faithfully infer the identity of the
object, ($\theta \in \{ \text{penguin} , \text{banana} \}$).

In practice, likelihoods often do not factorize exclusively through some
$\omega$ that can be measured in calibration experiments, but they depend on a
variety of additional stochastic latents $\omegap$.
In such cases, it is often not possible to derive the full dependence
$p(x|\omega, \omegap)$, but it may still be possible to correct the explicit
dependence on $\omega$ and remove the likelihood ``gap'' between reality and the
forward model used for inference.

\subsection{Calibration methods}
\label{sec:calib}

We are aware of two procedures for deriving these corrections to the simulated
density, $\psim(x|\omega)$, with calibration data drawn from $\pdata(x|\omega)$.
The first constructs the density ratio
\begin{equation}
   s_\omega(x) \equiv \frac{\pdata^\CR(x|\omega)}{\psim^\CR(x|\omega)}
\end{equation}
using observations in a ``calibration region'' (\CR).
Inference on $\theta$ is then performed on observations in the ``signal region''
(SR) using the corrected likelihood
\begin{equation}
   p(x|\theta) =
      \int d\omega\, d\omegap\,
         s_\omega(x)\, \psim(x|\omega, \omegap)\,
         \psim(\omega, \omegap | \theta).
\end{equation}
Thus, corrections derived in the CR are applied to the SR likelihood via the
weight $s_\omega(x)$; this is colloquially known as ``vertical calibration''.

An alternative strategy is to derive the conditional transport function
$T_\omega$ (and corresponding transport map $(T_\omega)_\#$) such that
\begin{equation}
   (T_\omega)_\#\, \psim^\CR(x | \omega)
   \equiv \psim^\CR(T_\omega\, x|\omega)
   = \pdata^\CR(x | \omega).
\end{equation}
This transport is subsequently applied in the SR, resulting in the corrected
likelihood 
\begin{equation}
   p(x|\theta) =
      \int d\omega\, d\omegap\,
         (T_\omega)_\#\, \psim(x|\omega, \omegap)\,
         \psim(\omega, \omegap | \theta).
\end{equation}
This ``horizontal calibration'' moves the simulated observations based on what
is observed in the calibration region.
The suitable transport map for correcting simulations is the optimal one for a
given choice of distance metric, i.e.~the one which minimally alters
$\psim(x|\omega)$ for each choice of $\omega$ while still closing with
$\pdata(x|\omega)$~\cite{Pollard_2022}. 
The optimal transport map is the solution to this closure problem that is least
invasive, in the sense that it preserves the full structure of $\psim(x|\omega,
\omegap)$ as well as possible. 

It is worth noting that, in the limit that $\omega$ captures all the mismodeling
of the simulation $\psim(x|\omega, \omegap)$, then the corrected simulation
converges to the ``real'' likelihood $\pdata$.
This is true for both vertical and horizontal strategies.
While it is not the focus of this article, the choice of $\omega$ against which
one parameterizes the corrections is important; enabling corrections that are
conditioned on a sufficiently informative $\omega$, of potentially very
high-dimensions, is an active field of research~\cite{ATLAS:2022zkl}.

Historically, vertical calibrations have usually been derived in discrete,
low-dimensional binnings, and horizontal calibrations are derived using
tractable likelihood ans\" atze; see
Refs.~\cite{ATLAS:2019EPJC79-970,ATLAS:2022jjr} for recent examples in collider
physics.
For example, if the predicted and true likelihoods are univariate Gaussian with
respective means and variances $(\mu_\mathrm{sim}, \sigma_\mathrm{sim}^2)$ and
$(\mu_\mathrm{data}, \sigma_\mathrm{data}^2)$, then horizontal calibration
amounts to applying the transformation
$x \mapsto \frac{\sigma_\mathrm{data}}{\sigma_\mathrm{sim}}
(x - \mu_\mathrm{sim} + \mu_\mathrm{data})$
to simulated samples.
These conventional methods are, however, limited to low-dimensional settings
and simple likelihood ans\" atze, largely due to data scarcity.

In the past several years, ML-based solutions to both the vertical and
horizontal calibration problems have been developed.
Density ratio estimation (DRE) via neural-network (NN) classifiers is a
well-established technique that scales well to high dimensions in both $x$ and 
$\omega$~\cite{Sugiyama_Suzuki_Kanamori_2012}. 
More recently, neural OT solvers have been
introduced~\cite{Original_ot_solution,cpflows,DBLP:journals/corr/abs-2106-01954,bunne2023supervisedtrainingconditionalmonge},
enabling the derivation of OT maps for horizontal calibrations of
multi-dimensional simulations from which samples can be
drawn~\cite{Pollard_2022, ATLAS-PAPER-2025}. 

As we scale into higher dimensions, it is important to recognize that the
horizontal and vertical strategies are distinct solutions to the
misspecification problem and exhibit different behaviors.
In sample-based simulations, the reweighting function $s(x)$ used in vertical
calibration will, in general, dilute the statistical power of the simulated
samples by a factor
\begin{align}
   \rho
      \equiv \frac{\mathbb{E}_{x\sim\psim}[s(x)^2]}{\mathbb{E}_{x\sim\psim}[s(x)]^2}
      = \mathbb{E}_{x\sim\pdata}[s(x)],
\end{align}
i.e.~the effective number of samples available after calibration is
$\Neff \equiv \Nsim/\rho$, where \Nsim is the original number of samples.
Appendix~\ref{sec:dilution} shows the derivation of this equality in more detail.
Horizontal calibration does not introduce a dilution.
For a more thorough treatment of these and other issues with reweighting- and
importance-sampling methods, we recommend Chapter~9 of Owen~\cite{mcbook}.

Indeed, for well-behaved spaces (e.g. Euclidean space), a horizontal solution to
a given calibration problem always exists~\cite{villani2008optimal}.
For vertical calibrations, a solution only exists under certain conditions; at
the very least,
$\forall\, x\ \omega.\ \pdata(x|\omega) > 0 \Rightarrow \psim(x|\omega) > 0$
must hold, i.e. the simulated distribution must fully support the data
distribution such that $s_\omega$ is always finite.
However, as shown in Appendix~\ref{sec:dilution}, this is not in general
sufficient to guarantee existence of a vertical-calibration solution\footnote{
For a more in-depth discussion of importance sampling weights and their
behaviors in a variety of situations, including potential pitfalls, we recommend
Chapter~9 of Owen's \emph{Monte Carlo: Theory, Methods and
Examples}~\cite{Owen2013MonteCarlo} or Chapter~3 from Robert \&
Casella~\cite{RobertCasella2004}.}.
These properties originate from the fact that an OT \emph{distance} between
densities $p$ and $q$ over a space $Y$ is well-defined under rather loose
conditions on $Y$~\cite{villani2008optimal}, while vertical calibrations
inherently depend on the $\chi^2$-\emph{divergence} between $p$ and $q$ to be
finite~\cite{Rényi1961Entropy,Csiszár1967InformationType}.

As derived in Appendix~\ref{sec:dilution}, even in simple scenarios in which
$\pdata$ and $\psim$ are approximately multivariate-normal distributed (MVN),
there are stringent requirements on how overlapping they must be for vertical
calibrations to be viable.
In the MVN case, if the variance of $\pdata$ is at least twice that of $\psim$
in even one dimension, then the dilution factor is infinite, and vertical
calibrations are not possible.
Even if the covariances are identical, the dilution factor grows exponentially
with the Mahalanobis distance~\cite{Mahalanobis2018Reprint} of the means of
$\pdata$ and $\psim$, again limiting the utility of the vertical strategy.
Indeed, in the particle physics-inspired example explored in
Section~\ref{sec:results}, we conclude that $\rho$ is very large, and a vertical
calibration solution may not exist.

\subsection{Hierarchical models in collider physics}

In this article we explore a concrete example of an important calibration
problem in particle physics.
Data analysis in High Energy Physics (HEP) experiments at the Large Hadron
Collider (LHC) aims to study the fundamental properties of matter by analyzing
the complex final states resulting from high-energy
collisions~\cite{Evans:2008zzb}.
These analyses rely on hierarchical stochastic calculations, where simulations
of the underlying ``hard process,'' involving fundamental particles,
approximately factorize from the subsequent evolution into stable particles that
eventually interact with the detector.
These detector signals undergo sequential processing through a series of
algorithms defining the reconstruction chain of the
experiment~\cite{Buckley:2011ms,ATLAS:2010arf,GEANT4:2002zbu,Allison:2006ve,Allison:2016lfl}.
Ultimately, dedicated analyses are conducted to extract key parameters,
$\theta$, of the underlying theoretical models of interest that govern the
initial hard process.

Within this framework, Monte Carlo (MC) simulations play a crucial role in
modeling the likelihood via sampling.
Although MC simulations are grounded in first-principles, they often exhibit
substantial discrepancies when compared to experimental data, rendering
$\psim(x|\omega,\theta)$ unreliable at the precision required for inference.
Given that these likelihoods are inherently intractable, controlling every
potential source of mismodeling from first principles is impractical.  
Instead, dedicated calibration regions are defined and corrections derived.
The conditioning variables $\omega$ often comprise kinematic properties related
to the reconstructed observable $x$; the transverse momentum of a jet with
respect to the incoming proton beams ($p_\mathrm{T}$) and its pseudorapidity
($\eta$) are common choices for conditioning quantities.

\subsection{Hadronic jets and their representations}

Here we focus on an important use-case that is ubiquitous across the LHC physics
programme: characterizing jets of hadronic particles (``jets'').
For example, establishing what was the origin of a jet observed in a detector is
critical for a broad number of important conclusions drawn from LHC data.
ML techniques have vastly improved our ability to answer this question to high
precision in the last few years~\cite{UParT,GN2}.

Even more recently, foundation models have emerged as promising tools for
enhancing reconstruction tasks, summarizing the relevant features of jets in a
general way~\cite{MPM,sophon,Mikuni,Vigl2024,Birk_2024}.
Foundation models used for jet tagging are essentially large-scale
transformer-based networks that can be pre-trained to encode high-dimensional
representations of jets, $z$, but maintain nearly all relevant information
contained in the original jet constituents.
These general, powerful latent representations may subsequently be used as
inputs for a large body of downstream inference tasks.
Indeed, the number of measurements from the LHC experiments that depend on jet
reconstruction and identification each year is measured in the dozens, but each
of these uses different subsets of the jet information content to constrain
underlying physical parameters.

While it's possible to derive less-general but still-optimal jet representations
for each individual analysis relying on jet information, foundation models offer
significant potential for enhancing HEP reconstruction pipelines exactly because
they provide general-purpose representations that can be used for a broad range
of measurements.
However, the experimental methodology for calibrating discrepancies between MC
simulations and real experimental data for these representations, which are of
high dimensionality, is currently lacking.
A properly-calibrated, universal data representation enables unbiased inference
for a wealth of downstream tasks and is critical for expanding the reach of
the LHC and similar experiments.
Indeed, by calibrating a rich, informative jet representation, the
hundreds of downstream analyses benefit without needing to individually perform
their own calibrations, which is not feasible at this
scale~\cite{GN2,ATLAS:2022qxm}.

In this work, we build upon previous research and, for the first time, introduce
a calibration approach leveraging OT theory, applied directly in a
high-dimensional latent space of up to 128 dimensions.
Calibration algorithms based on optimal transport have been introduced in
Ref.~\cite{Pollard_2022,ATLAS-PAPER-2025} and applied to correct
low-dimensional data representations, such as the output scores of classifiers
used for jet tagging in Ref.~\cite{GN2,ATLAS:2019EPJC79-970}.
We demonstrate this methodology's promise in addressing misspecification for a
much more informative set of features from the latent representation of a
general-purpose jet tagging classifier trained over a wide variety of particle
jets.
While we concentrate on the objective of jet tagging, the applicability of this
technique extends to a wide range of reconstruction tasks in HEP experiments and
beyond.

    \section{Related work} \label{sec:related_work}
    Optimal transport has recently emerged as a powerful tool for constructing high dimensional maps between distributions, with a growing body of applications in both theoretical and experimental physics~\cite{Frisch_2002,levy2022monge,vonhausegger2021bao,otsham2024}. 

In high-energy physics (HEP), one of the primary applications of OT is to correct discrepancies in simulated datasets through calibration~\cite{Pollard_2022}. 
This approach has been adopted by the ATLAS experiment, which employed OT to calibrate algorithms used for $b$-jet identification~\cite{ATLAS-PAPER-2025}. 
To the best of the author's knowledge, this represents the first instance of a horizontal calibration method applied to jet tagging. 
In contrast, conventional jet-tagging calibration techniques predominantly rely
on vertical calibration
strategies~\cite{ATLAS:2019EPJC79-970,CMS:2020P06005,ATLAS-PHYS-PUB-2021-035,ATLAS-CONF-2017-064,CMS:2022P03014}.
Vertical calibration strategies extend beyond jet-tagging, representing a standard procedure for various reconstructed objects in HEP experiments, including muons~\cite{ATLAS:Muon,CMS:Muon}, 
electrons and photons~\cite{ATLAS:egamma,CMS:egamma}, as well as tau leptons~\cite{ATLAS:Tau,CMS:Tau}.

Alternative horizontal calibration methods have been proposed, e.g.~via
normalizing flows~\cite{Daumann:2024kfd}, but it is important to recognize that
these solutions are known not to converge on the optimal-transport map in
real-world situations~\cite{Algren_2024}, nor are the solutions unique. 
By contrast, optimal-transport–based calibration admits a solution that is guaranteed to exist and be unique, 
as discussed further in Appendix~\ref{sec:optimal_transport}.

OT has also been explored in the modeling of complex background processes in the context of double Higgs bosons production at the LHC~\cite{manole2022background}, for anomaly detection algorithms~\cite{craig2024anomaly}, and for variable decorrelation~\cite{Algren_2024}.

%Beyond HEP, OT techniques have found significant applications in cosmology. 
%OT methods have been used to map the present-day galaxy distribution back to primordial density fields~\cite{Frisch_2002,levy2022monge}. In the context of Baryon Acoustic Oscillation measurements, 
%OT has been shown to enhance the reconstruction of the linear density field from low-redshift data~\cite{vonhausegger2021bao}. 
%Furthermore, OT has been explored as a tool for subhalo abundance matching to model the galaxy-halo connection~\cite{otsham2024}.

These advances are grounded in foundational developments by Monge and Kantorovich~\cite{villani2008optimal}, as well as in algorithmic contributions that enabled scalable applications of OT in machine learning~\cite{peyre2019computational}. 
Beyond applications in the physical sciences, the versatility of OT algorithms is well-established in computer vision tasks such as style transfer. 
In this context, OT-based maps are employed to render the content of one image using the stylistic features of another~\cite{style_matter}. 
This is conceptually similar to the calibration problem addressed in this work, where we seek to $render$ simulated events to match the characteristics of experimental data while preserving their underlying physical information.
High-dimensional OT has also been explored in the context of modifying the latent space of an autoencoder~\cite{bunne2023supervisedtrainingconditionalmonge,Original_ot_solution}.
%Together, these works have established OT as a powerful tool for tasks involving high-dimensional calibration, variable decorrelation, background modeling, and sample-to-sample correspondence.

    \section{Simulated dataset} \label{sec:simulated_dataset}
    
The dataset utilized in this study is based on \texttt{JetClass}~\cite{JetClass}, a well-established benchmark for jet classification tasks in HEP. 
It is generated using standard Monte Carlo event generators widely adopted in LHC experiments and incorporates detector response and reconstruction effects simulated with \texttt{DELPHES}~\cite{delphes} (v3.4.2). 

Jet reconstruction is performed using the anti-$k_T$ clustering algorithm~\cite{Cacciari_2008,Cacciari_2012} with a radius parameter of $R = 0.8$, applied to the E-Flow objects reconstructed by \texttt{DELPHES}. 
The analysis focuses on jets with transverse momenta in the range of $500$–$1000$~GeV and pseudorapidity satisfying $|\eta| < 2$.

Each jet in the dataset is categorized into one of ten distinct classes ($l$), 
each reflecting the underlying physics process responsible for its production. 
The classes comprise Higgs boson decays ($H \rightarrow b\bar{b}$, $H \rightarrow c\bar{c}$, 
$H \rightarrow gg$, $H \rightarrow 4q$, $H \rightarrow l\nu qq$), top quark decays ($t \rightarrow bqq$, $t \rightarrow b\ell\nu$), 
hadronic decays of the $W$ and $Z$ bosons, and a background class denoted $q/g$, 
which corresponds to jets initiated by quarks or gluons not included in the previous categories: 

\begin{equation}
    l \in {Hbb, Hcc, Hgg, H4q, Hl\nu qq, tbqq, tbl\nu, Wqq, Zqq, q/g}.
    \label{eq:labels}
\end{equation}

This study focuses on the calibration of $q/g$ jets which are abundantly produced at the LHC. The samples are generated using QCD multijet processes simulated with \texttt{Pythia} v8.186~\cite{Sj_strand_2015}. Parton showering and hadronization are also modeled with the same generator.  
Additionally, multiple parton interactions are included in the simulation, while pile-up effects are not considered.

\begin{figure*}[htpb]
    \centering
    \subfloat[]{{\includegraphics[width=0.40\textwidth]{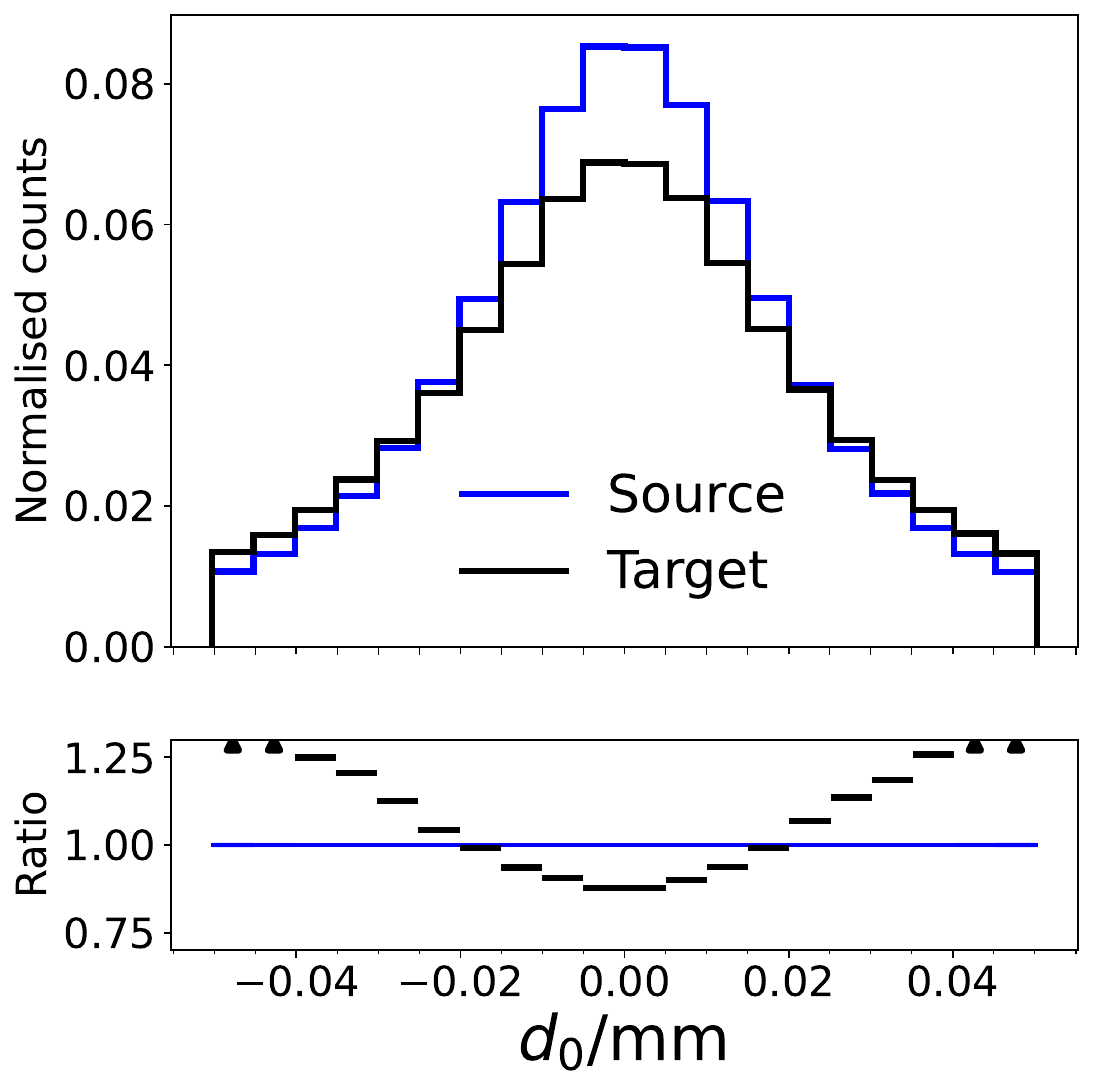}}}
    \subfloat[]{{\includegraphics[width=0.40\textwidth]{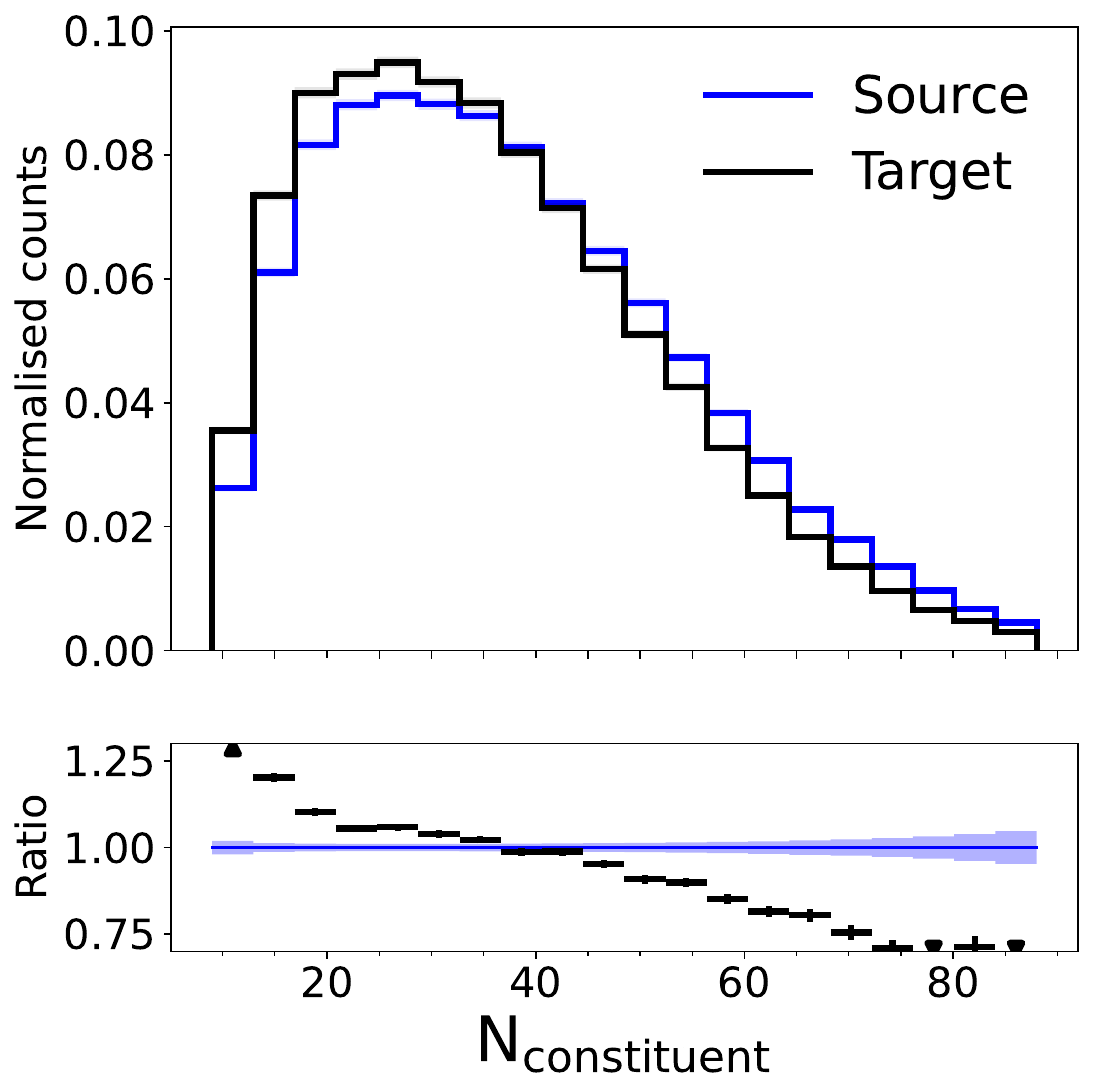}}}
    \caption{
        The distributions are represented as follows: source distribution (blue) and target distribution (black).
        (a) shows the change in the transverse impact parameter ($d_0$), while (b) illustrates the change in number of constituents ($N_{constituent}$) between the datasets.
    }
    \label{fig:data_variation}
\end{figure*}

To emulate the typical discrepancies observed between simulated samples and real collision data in LHC experiments, two datasets of QCD multijet events are generated. 
The first dataset, referred to as the \textit{source}, is produced using the standard \texttt{JetClass} configuration. The second dataset, referred to as the \textit{target}, 
is created by modifying the smearing functions of the \texttt{DELPHES} detector simulation. 
In particular, the impact parameters of charged particles are additionally smeared by 10\%--30\%, based on informed estimates from Ref.~\cite{ATLAS-CONF-2018-006}. 
In parallel, the renormalization and factorization scales of the QCD multijet events are increased from 1 to 2. 
These modifications are intended to simulate two main sources of systematic differences: inaccuracies in the modeling of the underlying physical processes, and effects stemming from detector response and reconstruction. 
Figure~\ref{fig:data_variation} compares the distributions of the transverse impact parameter ($d_0$) and the number of jet constituents for the two datasets. 
The implications of the discrepancies between the \textit{source} and \textit{target} datasets are discussed in detail in Section~\ref{sec:results}, where we demonstrate how these differences affect the physical observables resulting from jet classification.
%The production and decay of top quarks, as well as W, Z, and Higgs bosons, 
%are modeled with MadGraph5\_aMC@NLO~\cite{Alwall_2014} (v3.1.0). 
%The subsequent parton shower and hadronization are performed using  PYTHIA~\cite{Sj_strand_2015} (v8.243),
%which evolves the partonic system to produce the final outgoing particles. 

    \section{Method}
    \subsection{Constructing the classifier}

The identification of jet origin, commonly referred to as jet tagging, is a key challenge in particle physics, 
where machine learning techniques have become increasingly prominent. In this work, we construct a classifier based on a transformer~\cite{transformer} architecture that closely resembles the state-of-the-art models 
currently employed for jet-tagging at LHC experiments~\cite{GN2,UParT}.  

The classifier is designed to estimate the conditional probability distribution for the origin of a jet, 
$p(l \mid \{\vec{c}_{\text{obs}}\}, \text{jet}_{\text{obs}})$, 
where $\vec{c}_{\text{obs}}$ denotes the set of features associated with the jet constituents, and 
$\text{jet}_{\text{obs}}$ is a fixed-size vector of jet-level observables, including the pseudorapidity $\eta$, azimuthal angle $\phi$, transverse momentum $p_T$, energy $E$, and constituent multiplicity $N$.
A complete list of input variables, along with their corresponding distributions, is provided in Appendix~\ref{sec:jetfeat}.

The classifier uses a dedicated \textit{ClassToken}~\cite{cls_token} to aggregate the input set $\vec{c}_{\text{obs}}$ into a fixed 512-dimensional latent representation.
The Transformer encoder is composed of 4 encoder layers, each consisting of a multi-head self-attention mechanism followed by a feed-forward neural network~\cite{transformer}.
The jet input features $\text{jet}_{\text{obs}}$ are integrated into the encoder via Feature-wise Linear Modulation (FiLM), following the methodology established in Ref.~\cite{FiLM_newer}.

\begin{figure*}[htpb]
    \centering
    \subfloat[]{{\includegraphics[width=0.65\textwidth]{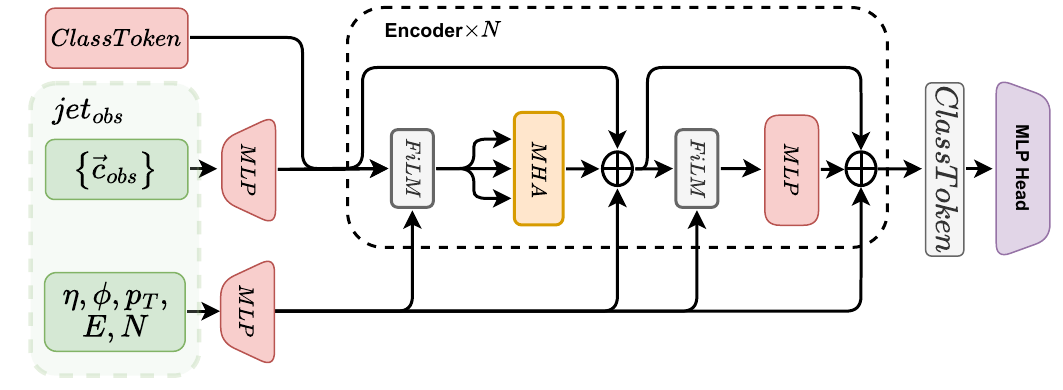}}}
    \subfloat[]{\includegraphics[width=0.25\textwidth]{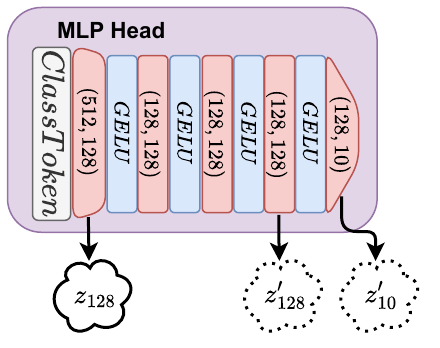}}
    \caption{
    The two diagrams show (a) the transformer encoder and (b) the output multi-layer perceptron head (\OMLP).
    The transformer architecture incorporates a sequence of encoder layers~\cite{prelayer_norm}, each comprising 
    multi-head self-attention mechanisms (MHA)~\cite{transformer} followed by feed-forward neural networks (MLP)~\cite{shazeer2020gluvariantsimprovetransformer} 
    with Feature-wise Linear Modulation (FiLM)~\cite{FiLM,FiLM_newer} for conditional integration.
    The \OMLP consists of a sequence of linear transformations and GELU~\cite{gelu} with a final output size of $10$ for the class probabilities.
    }
    \label{fig:transformer_architecture}
\end{figure*}

The jet's latent representation is passed to the output multi-layer perceptron head (\OMLP), which consists of a sequence of linear layers with GELU activation functions~\cite{gelu}, each maintaining a fixed latent dimensionality of 128. 
The final output of the network is a 10-dimensional latent vector containing the unnormalized class scores (logits) corresponding to the different jet categories. An overview of the classifier architecture is shown in \cref{fig:transformer_architecture}. 
The first 128-dimensional layer of the \OMLP, denoted $z_{128}$, is the target of the calibration. 
The final 128-dimensional layer ($z'_{128}$) and the output layer ($z'_{10}$) are also shown in \cref{fig:transformer_architecture} with dashed lines, and are used in the evaluation of the calibration procedure, as further discussed in Section~\ref{sec:results}.

The model is trained using the publicly available \texttt{JetClass} dataset~\cite{JetClass}. Details of the training procedure are provided in Appendix~\ref{sec:hyperparameters}.

\subsection{Deriving the optimal transport map}

The primary objective of this work is to derive an optimal transport map, denoted by $\widehat{T}_\omega$~\cite{Kantorovich2006OnTT,monge1781memoire}, such that
\begin{equation}
(\widehat{T}_\omega)_\#\, p_{\text{sim}}(z_{128} \mid \omega) = p_{\text{data}}(z_{128} \mid \omega),
\end{equation}
where $p_{\text{sim}}$ and $p_{\text{data}}$ represent the distributions of the latent representation $z_{128}$ in simulation and real data, respectively, conditioned on $\omega$. 
The conditioning variable $\omega$ typically includes jet kinematic features such as jet-$p_T$, jet-$\eta$, and other relevant labels extracted from the simulated dataset~\cite{ATLAS-PAPER-2025}. 
In this work we neglect dependencies on the jet kinematics and thus restrict $\omega$ to the $q/g$ label of the simulated sample that we aim to calibrate.

The optimal transport map $\widehat{T}_\omega$ is a unique transformation that minimally distorts 
\psim in order to match \pdata~\cite{Kantorovich2006OnTT,monge1781memoire,villani2008optimal}. Refs.~\cite{Original_ot_solution, cpflows, Algren_2024, ATLAS-PAPER-2025, DBLP:journals/corr/abs-2106-01954, bunne2023supervisedtrainingconditionalmonge} 
have shown that $\widehat{T}_\omega$ can be learned by minimizing the Euclidean transport cost between the simulated and observed data distributions, 
using a pair of input-convex neural networks (ICNNs)~\cite{ICNN,cpflows}. 
This approach is formalized by the following optimization problem\cite{villani2008optimal}:
\begin{equation}
    W_2^2(z, y) = \sup_{f(y)\in \text{cvx}(y)} \inf_{g(z)\in \text{cvx}(z)} f(\nabla_z g (z)) - \langle z, \nabla_z g(z) \rangle - f(y),
    \label{eq:OT_loss_function}
\end{equation}
where $f$ and $g$ are two distinct input-convex neural networks (ICNNs), $z \sim p_{\text{sim}}$ and $y \sim p_{\text{data}}$ denote samples from the simulated and target distributions, respectively. 
For simplicity, we have omitted the subscript ${128}$ from $z$ and $y$.

At convergence, the optimal transport map is given by $\widehat{T}_\omega(z|\omega) = \nabla_z g(z|\omega)$, 
which maps simulated latent representations $z$ to their calibrated counterparts
$y$, conditioned on
$\omega$~\cite{bunne2023supervisedtrainingconditionalmonge,ATLAS-PAPER-2025,Algren_2024}.
The derivation of the optimal transport map requires about 48 hours on a
mid-range consumer GPU with 8~GB of memory.
Both training and inference computational costs are small compared to the
simulation budget for contemporary collider experiments, which require dozens of
seconds to simulate each event~\cite{ATLAS:2021pzo}.

Further details on the ICNN architecture and training procedure are provided in Appendices~\ref{sec:optimal_transport} and~\ref{sec:hyperparameters}.

    \section{Results} \label{sec:results}

    % trained classifier
    
\subsection{Evaluating closure in the latent space} \label{sec:evaluating_closure}

%To evaluate the closure of the latent space calibration, a range of metrics will be employed. 
%The analysis begins with an exploration of the calibrated hidden representation in the $z_{128}$ space. 
%To quantify the agreement between the calibration and target distributions within this high-dimensional space, we introduce a discriminator, 
%denoted as \DL, trained to distinguish between the two.

%Finally, the calibrated hidden representations are passed through the subsequent frozen layers of the classifier, which transform them into output probabilities. 
%These probabilities are then examined to evaluate the extent of closure achieved by the calibration process in the output space.

    % 128d dimensional calibration
    
\begin{figure*}[htpb]
    \centering
    \includegraphics[width=0.48\textwidth]{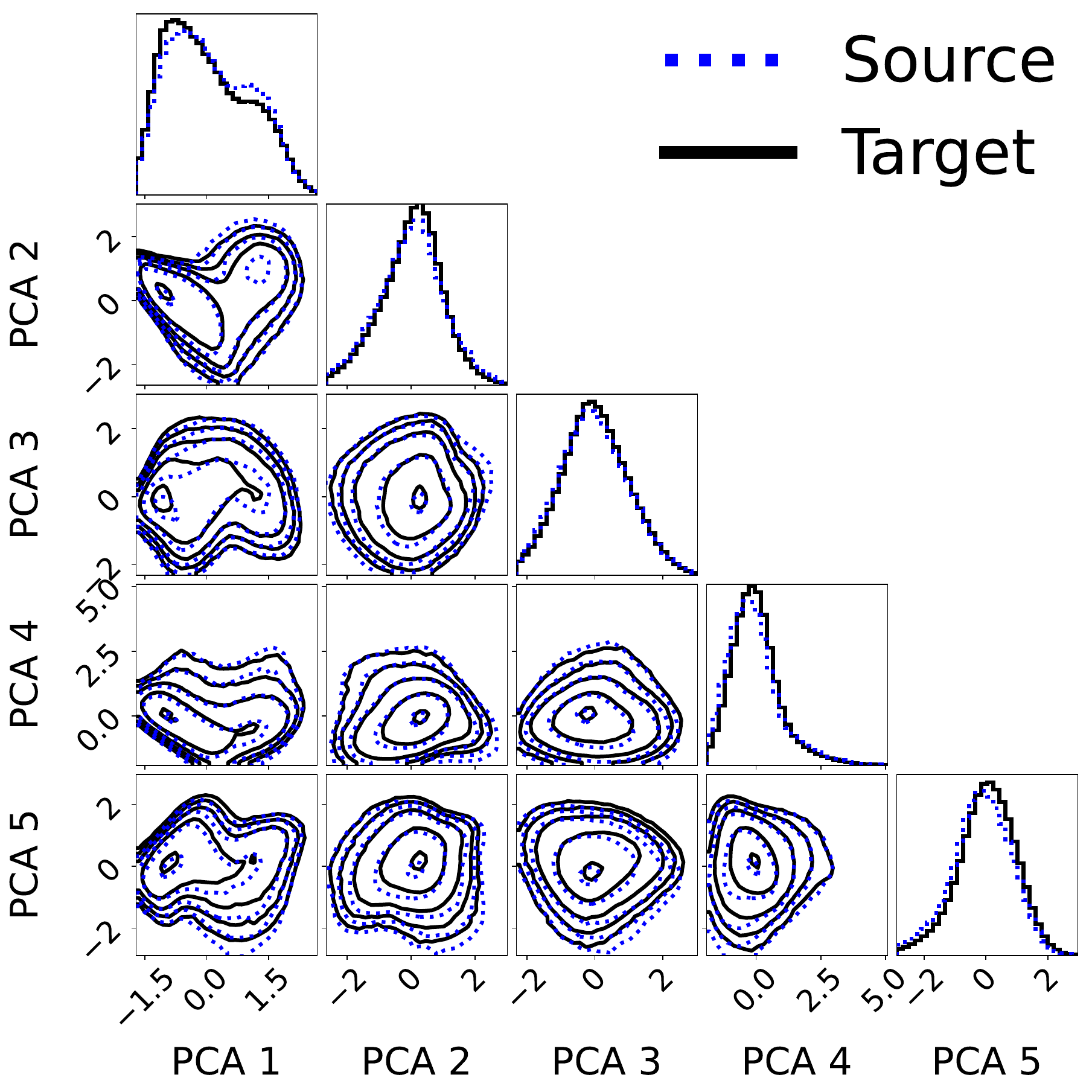}
    \includegraphics[width=0.48\textwidth]{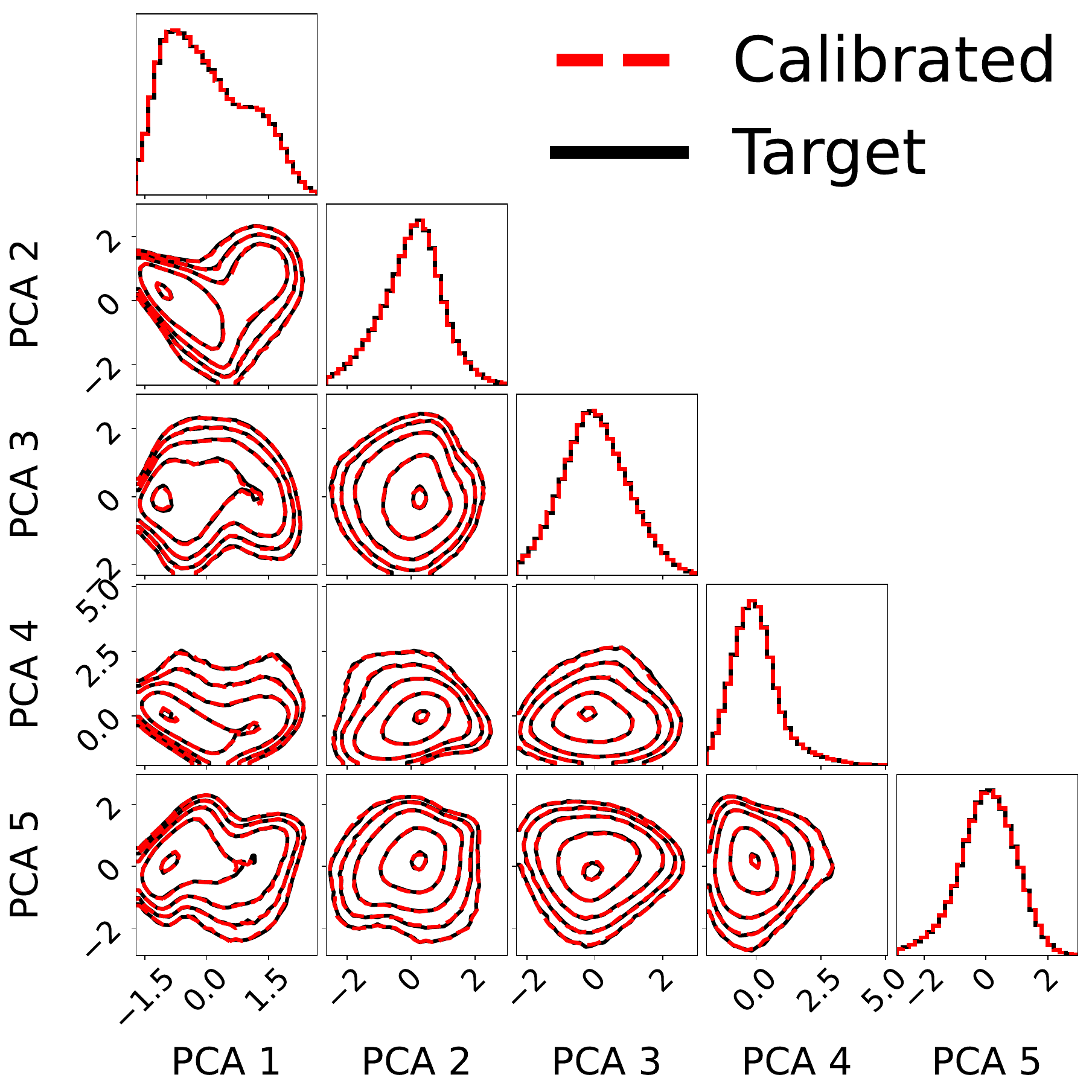}
    \caption{
        Corner plots of the first five principal components derived from PCA applied to the $z_{128}$ latent space. The 2D contours show 5\%, 50\%, 80\%, 90\%, and 95\% percentiles. 
        (a) Comparison between the target distribution (black) and the source distribution (blue). 
        (b) Comparison between the target distribution (black) and the calibrated distribution (red). 
        The calibration is derived in $z_{128}$, and subsequent layers in the \OMLP are applied to map it to the 10-dimensional output space.
        }
    \label{fig:pca_128_to_5}
\end{figure*}

To visualize the impact of calibration within the high-dimensional latent space $z_{128}$, 
Principal Component Analysis (PCA)~\cite{PCA} is applied to reduce the dimensionality to the five leading principal components. 
The distributions of these components are presented as corner plots in \cref{fig:pca_128_to_5}.

\cref{fig:pca_128_to_5}(a) illustrates the presence of the expected discrepancies between the source (simulated) and target (synthetic) data distributions. 
After applying the optimal transport map, \cref{fig:pca_128_to_5}(b) shows a significant reduction in these discrepancies, 
indicating successful alignment between the latent representations of the two domains.

\begin{figure*}
    \centering
    \subfloat[]{{\includegraphics[width=0.45\textwidth]{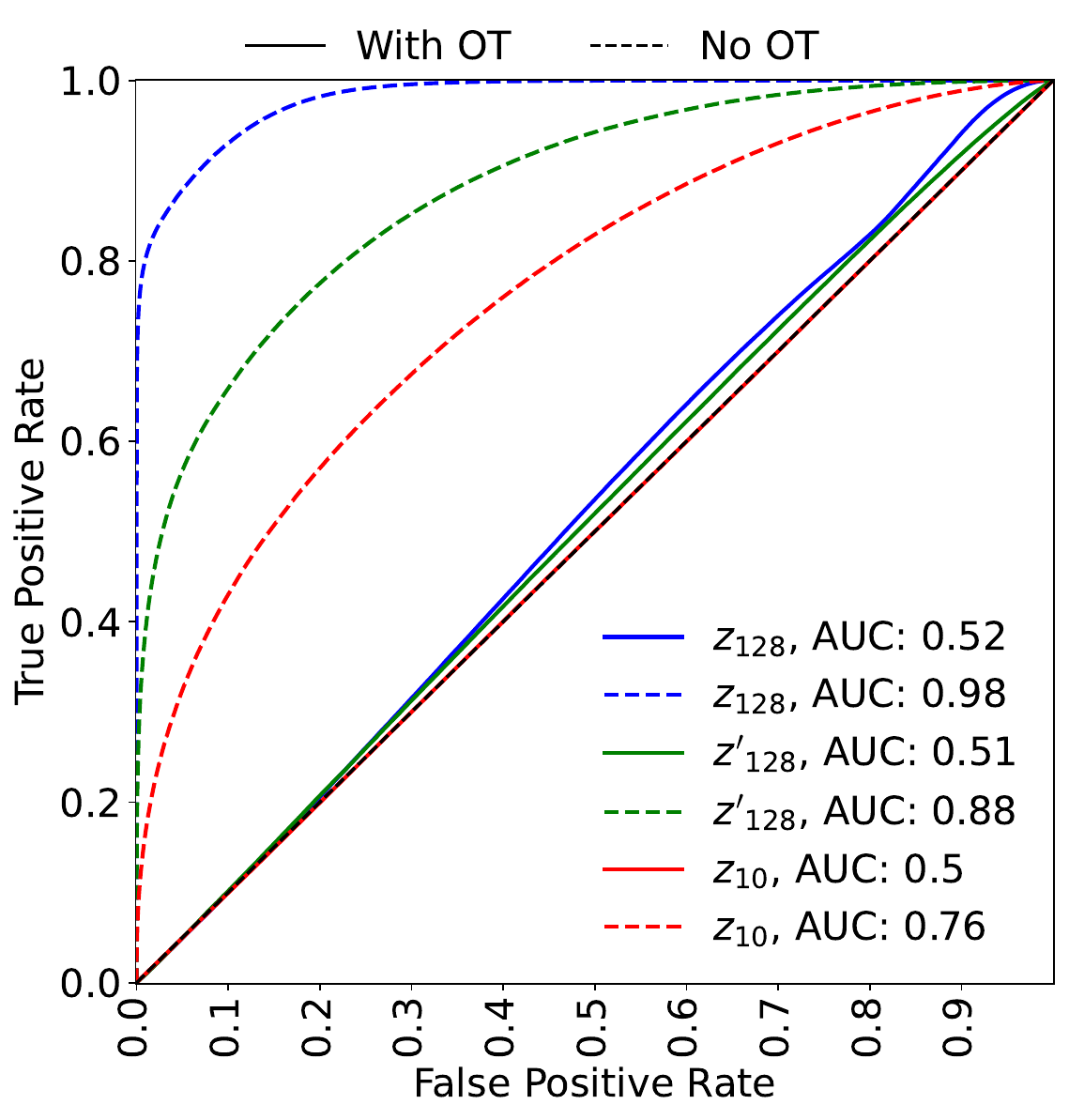}}}
    \subfloat[]{{\includegraphics[width=0.45\textwidth]{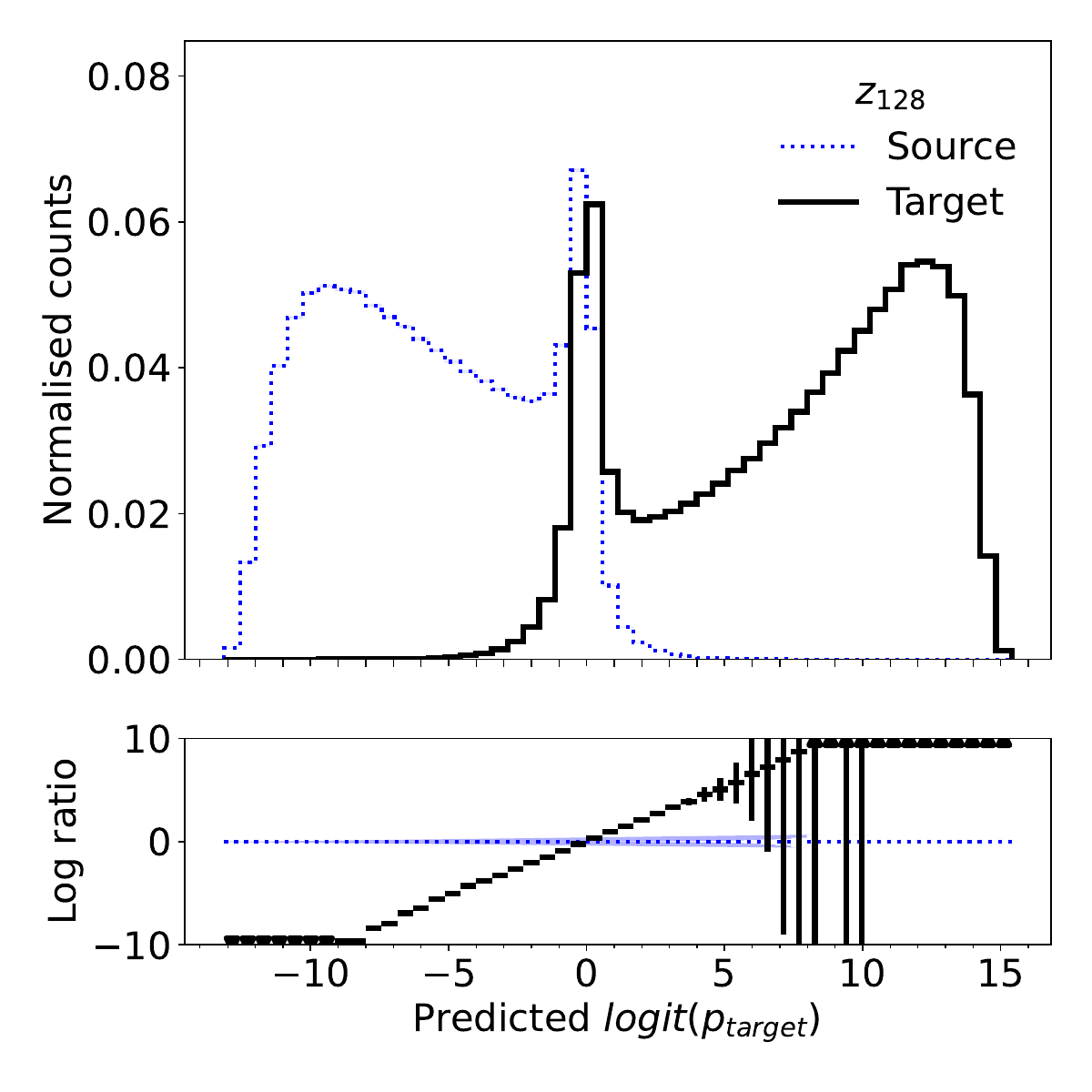}}}
    \caption{(a) Receiver operating characteristic (ROC) curves for discriminators \DL trained to differentiate between source and target distributions across three representational spaces: the 128-dimensional latent space ($z_{128})$, the final 128-dimensional layer in the MLP Head ($z'_{128}$), and the 10-dimensional output space ($z'_{10}$).
    The discriminators are subsequently evaluated on the calibrated versus target distributions, 
    where an area under curve (AUC) value approximating 0.5 indicates successful mitigation of distributional discrepancies that were previously exploited by the discriminator.
    (b) Marginal distribution of the discriminators \DL scores in $z_{128}$, between the source (blue) and target (black) distributions. 
    }
    \label{fig:only_128d_rocs}
\end{figure*}

The calibration procedure is further evaluated using discriminators, denoted \DL, trained to distinguish between the source and target distributions across  
the initial 128-dimensional layer ($z_{128}$), the final 128-dimensional layer ($z'_{128}$), and the output layer ($z'_{10}$) of the \OMLP. 
\cref{fig:only_128d_rocs}(a) presents the receiver operating characteristic (ROC) curves of \DL trained on each of these layers.
Discrepancies between the source and target distributions are largest in $z_{128}$ and decrease progressively in deeper layers, with the smallest differences observed in $z'_{10}$.

After applying the calibration to the source distribution, the discriminator ability to separate source from target is substantially diminished. 
A minor residual discrepancy persists in $z_{128}$ but vanishes in the deeper layers, resulting in area under the curve (AUC) values of 0.5 in the $z'_{128}$ and $z'_{10}$ latent spaces.

The discriminant score produced by \DL in the $z_{128}$ latent space is presented in \cref{fig:only_128d_rocs}(b). 
The source and target distributions exhibit only limited overlapping support and are otherwise nearly disjoint. 
In such cases, calibration methods based on reweighting—referred to as vertical calibration in section \cref{sec:calib} are subject to substantial statistical inefficiency due to the sparse overlap, 
thereby limiting their applicability. By contrast, the proposed horizontal calibration approach is not hindered by the lack of distributional overlap and thus remains a feasible strategy to calibrate high dimensional latent representations.

To further evaluate the performance of the calibration procedure, 
discriminators are trained to distinguish between the calibrated and target distributions within the latent spaces $z_{128}$, $z'_{128}$ and $z'_{10}$.
Notably, the discriminator achieves an AUC close to unity in the $z_{128}$ latent space as shown in~\cref{sec:ref2_comment}, 
indicating that either residual discrepancies persist post-calibration or new mismatches are introduced during the calibration process.
However, in the $z'_{128}$ layer, 
the AUC decreases to approximately 0.5, suggesting that these discrepancies are effectively attenuated by the subsequent transformations within the classifier. 
This result implies that residual discrepancies in the $z_{128}$ latent space do not propagate to the downstream layers and are therefore inconsequential for the final classification task.

\begin{figure*}[htpb]
    \subfloat[]{{\includegraphics[width=0.46\textwidth]{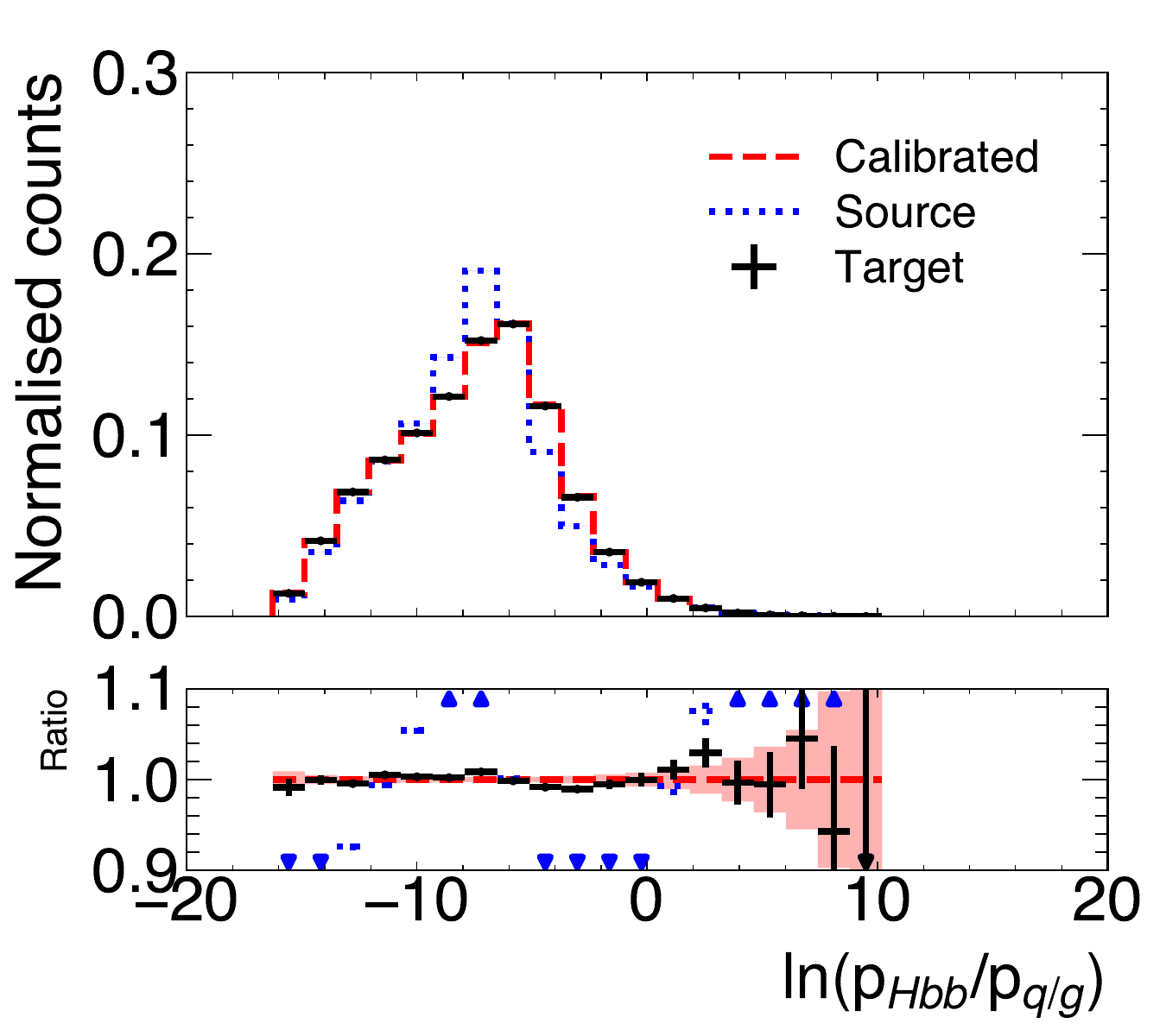}}}
    \hfill
    \subfloat[]{{\includegraphics[width=0.46\textwidth]{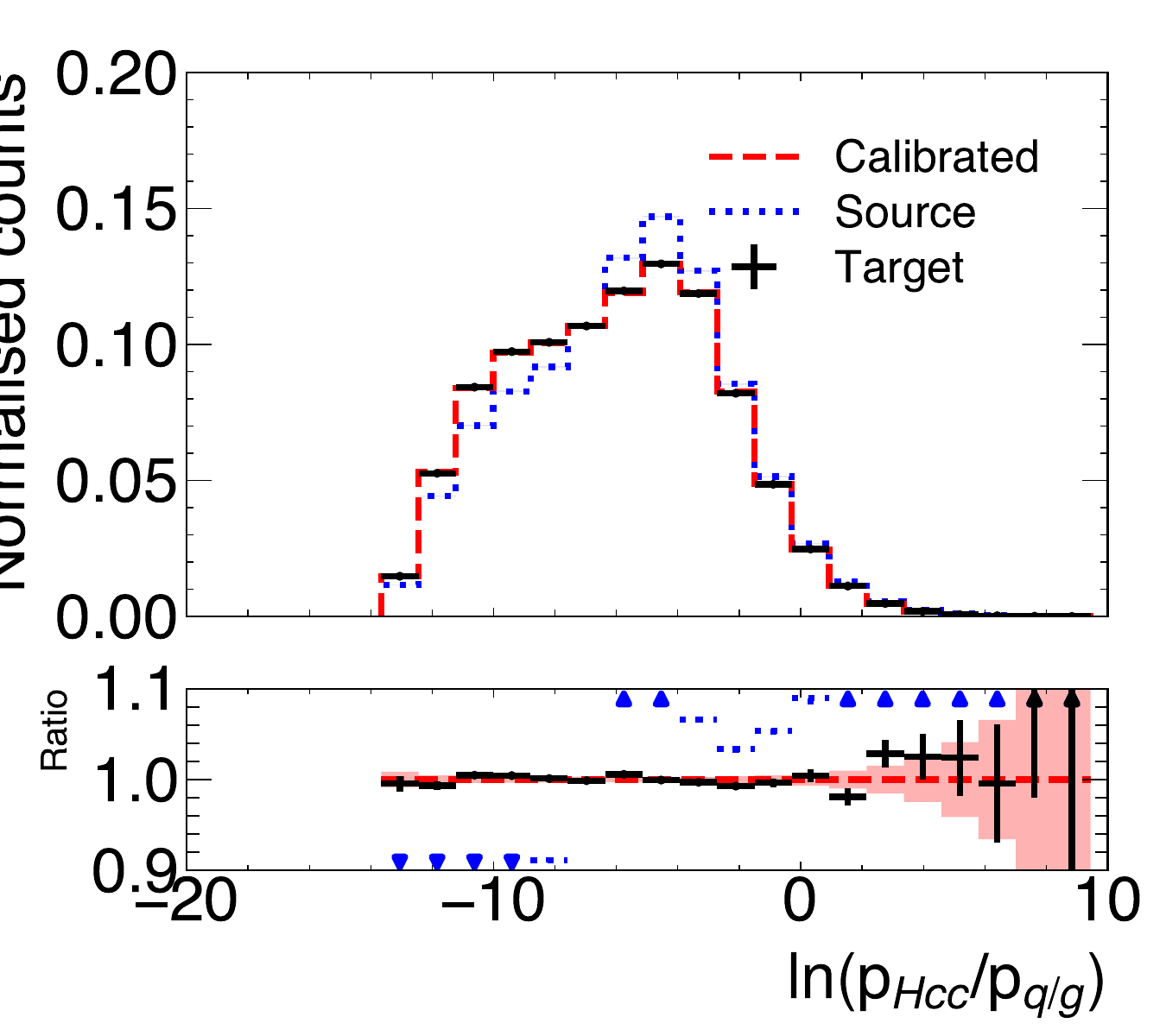}}}\\
    \centering
    \subfloat[]{{\includegraphics[width=0.46\textwidth]{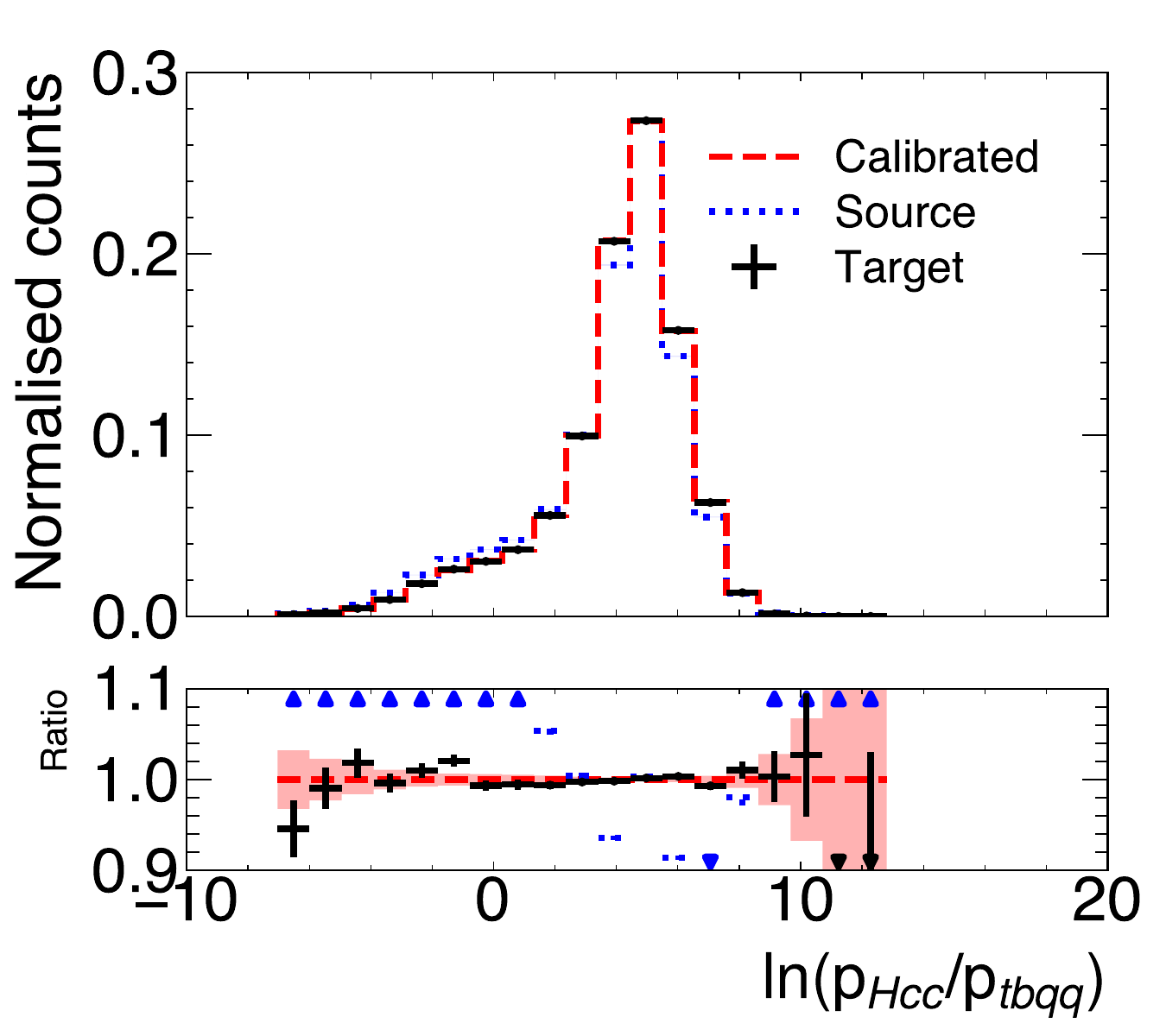}}}
    \caption{
        Comparisons between the marginal distributions of the physics-motivated one-dimensional discriminants.
        The distributions are represented as follows: target distribution (black), calibrated distribution (red), and source distribution (blue).
        The calibration is derived on the $z_{128}$ latent space, and then the subsequent \OMLP layers are applied to map it to the output space. Afterwards the output space is projected onto physics-motivated one-dimensional discriminants. 
        Description of the labels can be found in \cref{eq:labels}.
        }
    \label{fig:only_128d_marginal_distribution_physics_dics}
\end{figure*}

To validate this, the impact of the calibration is further examined in the classifier’s output space.
In typical applications output probabilities are combined to construct log-likelihood discriminants.
\cref{fig:only_128d_marginal_distribution_physics_dics} displays the distributions of these discriminants, where Higgs boson decays to $b$- or $c$-quark pairs are considered signal, 
while multijet and top-quark events define the background hypothesis.

Prior to calibration, clear discrepancies can be observed between the source (blue) and target (black) distributions.
The application of the calibration procedure (red) substantially reduces these differences, indicating improved agreement between the two distributions.

The jet flavour tagging calibrations discussed here, similar to more traditional calibration techniques, can be widely utilized across a large spectrum of physics analyses, such as measurements of Higgs bosons, Top quarks, and searches for new physics. Beyond jet flavour tagging, the methodology employed is pervasive in object reconstruction across particle physics. Consequently, although the degree of generalization depends on the specific case, the approach is fundamentally well-suited for broader implementation across other reconstruction tasks in the field

Technical details and supporting information for the methodology presented in 
this work are organized in the following Appendices: \cref{sec:dilution}
provides the formal derivation of the statistical dilution factor for vertical calibration 
with analytic solutions for the case of multi-dimensional gaussians; \cref{sec:optimal_transport} details the generalities of the optimal transport theory and the specific architecture of the Input-Convex Neural Networks (ICNNs); 
 \cref{sec:hyperparameters} describes the training procedures and the choice of hyperparameters of the classifiers; 
and \cref{sec:jetfeat} contains a comprehensive
 list of the input variables along with their respective distributions.

\clearpage

    % conclusion
    \section{Conclusion}

A calibration strategy based on optimal transport has been presented, 
targeting latent representations within transformer-based classifiers used in
high-energy particle physics.
By applying the calibration directly in the high-dimensional latent space, 
rather than on low-dimensional observables or classifier outputs, 
the proposed method enables more flexible and detailed correction of mismodeling
effects present in simulated data.

This methodology suggests a viable path toward the integration of foundation models in particle physics experiments. By adopting such an approach, it becomes possible to reduce the current reliance on numerous ad-hoc calibration analyses, thereby streamlining complex experimental pipelines.

Furthermore, this work invites a deeper exploration of calibration strategies, shifting the research focus from solely correcting model outputs toward the calibration of internal hidden representations and, potentially, input embeddings. Aligning these latent spaces offers a new level of architectural flexibility, providing greater control over feature representation and downstream task performance in particle physics applications.

Quantitative evaluations using discriminator networks indicate that the
calibrated latent representations can achieve good agreement with the target
domain for physics-relevant inference tasks.
While this work focuses on jet classification, the methodology is broadly
applicable to a range of reconstruction tasks in high-energy physics that
utilize deep learning models.

Further investigations into the non-closures observed in the calibration space are warranted, as ideally the calibration should not introduce new discrepancies.
Additionally, exploring finetuning strategies that leverage the calibrated latent
representations could enhance performance on downstream tasks.

Code is available at \url{https://github.com/malteal/LatentCalibration} and the datasets used in this study are available at \url{https://zenodo.org/records/18920733}.

    \section*{Acknowledgements}
    The MA and TG would like to acknowledge funding through the SNSF Sinergia
grant CRSII5\_193716 ``Robust Deep Density Models for High-Energy Particle
Physics and Solar Flare Analysis (RODEM)'' and the SNSF project grant
200020\_212127 ``At the two upgrade frontiers: machine learning and the ITk
Pixel detector''. 
CP acknowledges support through STFC consolidated grant ST/W000571/1.
\clearpage

    \appendix
    \section*{Appendix}
    \section{Reweighting dilution factors} \label{sec:dilution}

\subsection{General dilution factors}

The dilution factor $\rho$ (such that for $N$ samples from $p_1$, the effective
number of reweighted simulated samples is $\Neff \equiv \Nsim / \rho$) for $x
\sim p_1$ reweighted to $p_2$ via the ratio $s(x) \equiv p_2(x) / p_1(x)$ is
given by
$$ \frac{\mathbb E_{x\sim p_1} [s(x)^2]}{\mathbb E_{x\sim p_1} [s(x)] ^2}. $$
The denominator,
$$ \left( \int_x p_1(x) \frac{p_2(x)}{p_1(x)} \right) ^2, $$
is unity, since $p_2(x)$ is a probability density.
The numerator is given by
$$ \int_x p_1(x) \left( \frac{p_2(x)}{p_1(x)} \right)^2 = \int_x p_2(x) \frac{p_2(x)}{p_1(x)}, $$
i.e. the expected value over $p_2$ of $s$.
To summarize, the dilution factor is
$$ \rho = \mathbb{E}_{x\sim p_2}[s(x)] $$
and is related to the
$\chi^2$-divergence~\cite{Rényi1961Entropy,Csiszár1967InformationType}:
$\mathbb{E}_{x\sim p_2}[s(x)] = 1 + \chi^2(p_2||p_1)$.
The minimum $\rho$ is unity when $\forall x.~p_1(x) = p_2(x)$, i.e.~no
reweighting is required.

\subsection{Dilution factors for multivariate normal distributions}

To make the previous discussion concrete, we derive analytic expressions for
$\rho$ for two multivariate normal distributions (MVNs).
Similar examples and a more formal treatment may be found in Chapter~9 of
Owen~\cite{mcbook}; see e.g.~Example~9.1 therein.
The log-density ratio of two MVNs $p_1$ and $p_2$, with respective means $\mu_1,
\mu_2$ and covariances $\Sigma_1, \Sigma_2$, is given by
$$
2 \log\frac{p_2(x)}{p_1(x)}
  = \log\frac{|\Sigma_1|}{|\Sigma_2|}
    - (x-\mu_2)^T \Sigma_2^{-1} (x-\mu_2)
    + (x-\mu_1)^T \Sigma_1^{-1} (x-\mu_1).
$$

The dilution factor
$\rho = \mathbb{E}_{x\sim p_2}[s(x)] = \int_x p_2^2(x) / p_1(x)$
in this case is the  result of a Gaussian integral.
To ease notation, we define
$$
A \equiv 2\Sigma_2^{-1} - \Sigma_1^{-1}
  \qquad b \equiv 2\Sigma_2^{-1}\mu_2 - \Sigma_1^{-1}\mu_1
  \qquad c \equiv \mu_2^T \Sigma_2^{-1} \mu_2 - \frac 1 2 \mu_1^T \Sigma_1^{-1}\mu_1,
$$
yielding for $n$-dimensional MVNs
\begin{align*}
\rho
  & = \frac{\sqrt{|\Sigma_1|}}{(2\pi)^{n/2}|\Sigma_2|} \int_{\mathbb{R}^n}
      \exp\left( - \frac{1}{2} x^T A x + b^T x - c \right) \\
  & = \frac{\sqrt{|\Sigma_1|}}{|\Sigma_2|\sqrt{|A|}}
      \exp\left(\frac 1 2 b^T A^{-1} b - c \right).
\end{align*}

$\rho$ is only finite if $A$ is positive definite.
In case $\Sigma_1$ and $\Sigma_2$ are simultaneously diagonal, then the dilution
factor is infinite unless $\forall i.~A_{ii} > 0$, i.e.~in every dimension the
variance of $p_1$ can be no less than half that of $p_2$.
This prevents the vertical calibration procedure (and reweighting in general)
from being viable in a wide variety of common scenarios.

If $\Sigma_1 = \Sigma_2 = \Sigma$, then $\rho$ is simply $\exp (\Delta^T \Sigma
\Delta)$ with $\Delta \equiv \mu_1 - \mu_2$.
In this case the dilution factor grows exponentially with the Mahalanobis
distance between the means~\cite{Mahalanobis2018Reprint}.
If $\mu_1 = \mu_2 = \mu$, then
$$ \rho = \frac{1}{|\Sigma_2|} \sqrt{\frac{|\Sigma_1|}{|A|}}; $$
this depends only on the relative covariances and, as stated earlier, diverges
when $A$ is not positive definite.

    \section{Optimal transport with input-convex neural networks}
\label{sec:optimal_transport}

Optimal transport theory addresses the fundamental problem of determining an efficient 
mapping between probability distributions, 
specifically between source distribution $Q$ and target distribution $P$. 
Following the notation established in the introduction, 
we denote $Q=p_{\text{sim}}$ and $P=p_{\text{data}}$.
The concept of optimality is in relation to minimizing of a cost function $c(x,y)$, 
which quantifies the price of transporting density between $p_{\text{sim}}$ and $p_{\text{data}}$.

The classical primal formulation of the optimal transport problem~\cite{monge1781memoire}, is defined as:
\begin{equation}
    T^* = \inf_{T: T_{\#}p_{\text{sim}}=p_{\text{data}}} \frac{1}{2} \mathop{\mathbb{E}}_{Z\sim p_{\text{sim}}} \left[ \|z-T(z)\|^2 \right],
\end{equation}
where $T$ is the transport map and 
$\|\cdot\|$ is the cost function that ranks the maps. 
It is an optimization problem that seeks to find the map $T$ that minimizes $\|\cdot\|$ over all maps that satisfy $T_{\#}p_{\text{sim}}=p_{\text{data}}$, where $T_{\#}$ is the pushforward operator~\cite{villani2008optimal}.
Monge's transport maps are defined as deterministic maps,
which imposes significant analytical challenges~\cite{Kantorovich2006OnTT, villani2008optimal}. 
Consequently, Kantorovich proposed a relaxation of Monge's primal formulation by conceptualizing transport as a probability measure rather than a deterministic mapping~\cite{Kantorovich2006OnTT}. 
Additionally to the relaxation, 
Kantorovich introduced a dual problem that has the same solution as the primal one.
The dual problem is defined as:
\begin{equation}
    W_2^2(f,g)=\sup_{f(y)+g(z)\leq c(y,z)} \mathop{\mathbb{E}}_{Y\sim p_{\text{sim}}} \left[ f(y) \right] + \mathop{\mathbb{E}}_{Z\sim p_{\text{data}}} \left[ g(z) \right],
\end{equation}
where $f$ and $g$ are functions subject to the constraint $f(y)+g(z)\leq c(y,z)$~\cite{villani2008optimal}.
Ref.~\cite{Original_ot_solution} demonstrated 
that for continuous distributions, the optimal transport can be derived through the following optimization problem:
\begin{equation}
    W_2^2(y, z) = \sup_{f(y)\in \text{cvx}(y)} \inf_{g(z)\in \text{cvx}(z)} f(\nabla_z g (z)) - \langle z, \nabla_z g(z) \rangle - f(y),
    \label{eq:OT_loss_function}
\end{equation}
where $f(y)$ and $g(z)$ are convex functions over the respective domains.
The resulting optimal transport map is characterized as $T^*=\nabla_z g (z)$, which maps points from the source domain to the target domain.
\begin{figure*}[htpb]
    \centering
    \includegraphics[width=0.55\textwidth]{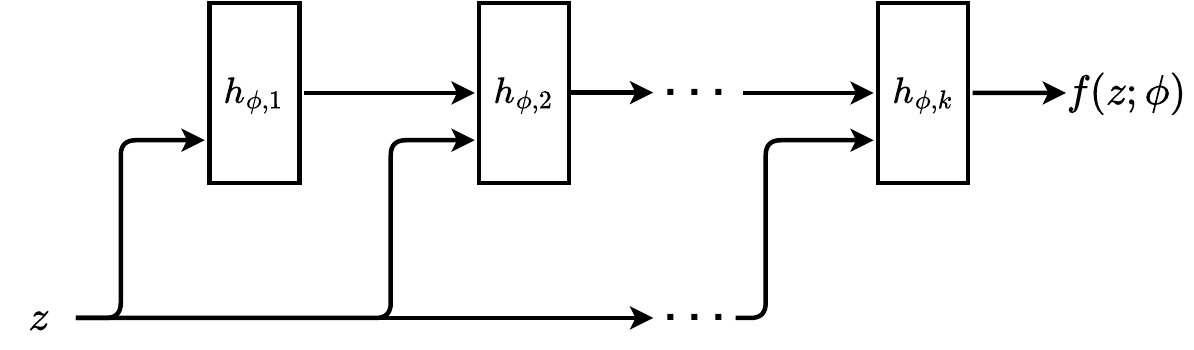}
    \caption{
    The diagram shows the architecture of the ICNN\cite{ICNN} used to derive the 
    optimal transport by optimizing \cref{eq:OT_loss_function}. 
    The specific implementation of the ICNN that is used in this paper follows the architecture described in~\cite{cpflows,Algren_2024}.
    }
    \label{fig:icnn_architecture}
\end{figure*}
\subsection{Method to derive the optimal transport} \label{sec:method_to_derive_optimal_transport}
As established in \cref{sec:optimal_transport}, 
the solution to \cref{eq:OT_loss_function} requires two convex functions $f$ and $g$.
While conventional neural networks (NN) are not limited to convex functions, 
Ref.~\cite{ICNN} introduced input-convex neural networks (ICNNs), 
which limit the parametrization of NNs to only convex functions
$f(y;\theta)$ with respect to their input variables $y$.
The ICNN architecture implements a recursive formulation across layers $i=0,...,k-1$, wherein the $i^{\text{th}}$ layer is defined as:
\begin{equation}
    \centering
    h_{i+1} = \sigma(W^h_i h_i + W^y_i y + b_i), \quad f(y;\theta) = h_k,
\end{equation}
where $h_0=W^h_i=0$, and $\sigma$ represents a convex, monotonically non-decreasing activation function.
A schematic representation of the ICNN architecture is illustrated in \cref{fig:icnn_architecture}.

There is an extensive literature on the ICNN and using it to derive the 
optimal transport~\cite{Original_ot_solution, cpflows,Algren_2024,ATLAS-PAPER-2025,DBLP:journals/corr/abs-2106-01954,bunne2023supervisedtrainingconditionalmonge}
with different tweaks to the ICNN architecture and the training procedure.

The specific architecture and training implementation of the ICNNs used in this paper follow the architecture of Ref.~\cite{cpflows} and training procedure of Ref.~\cite{Algren_2024}.

\begin{figure*}[htpb]
    \centering
    \subfloat[]{{\includegraphics[width=0.42\textwidth]{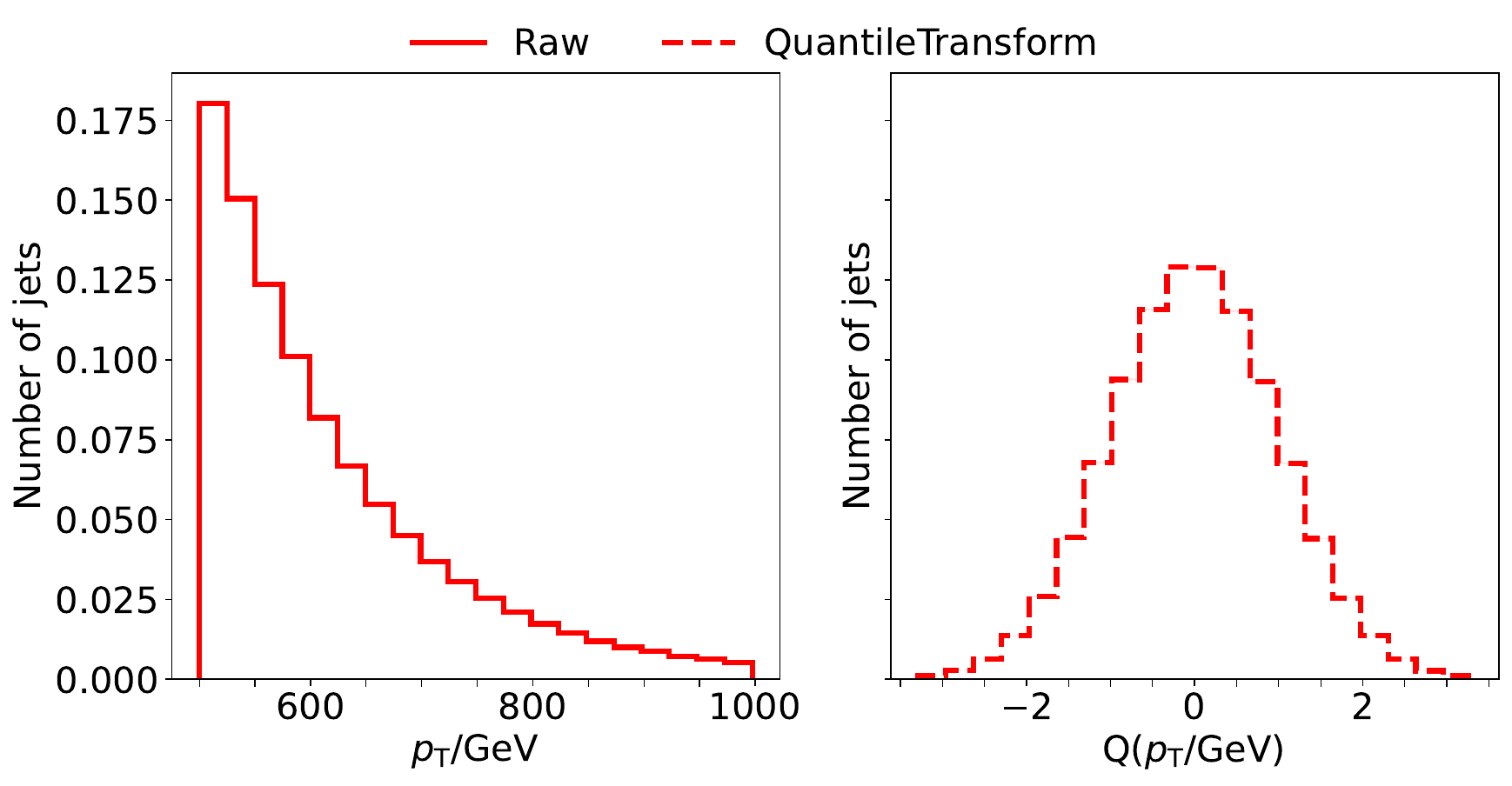}}}
    \quad
    \subfloat[]{{\includegraphics[width=0.42\textwidth]{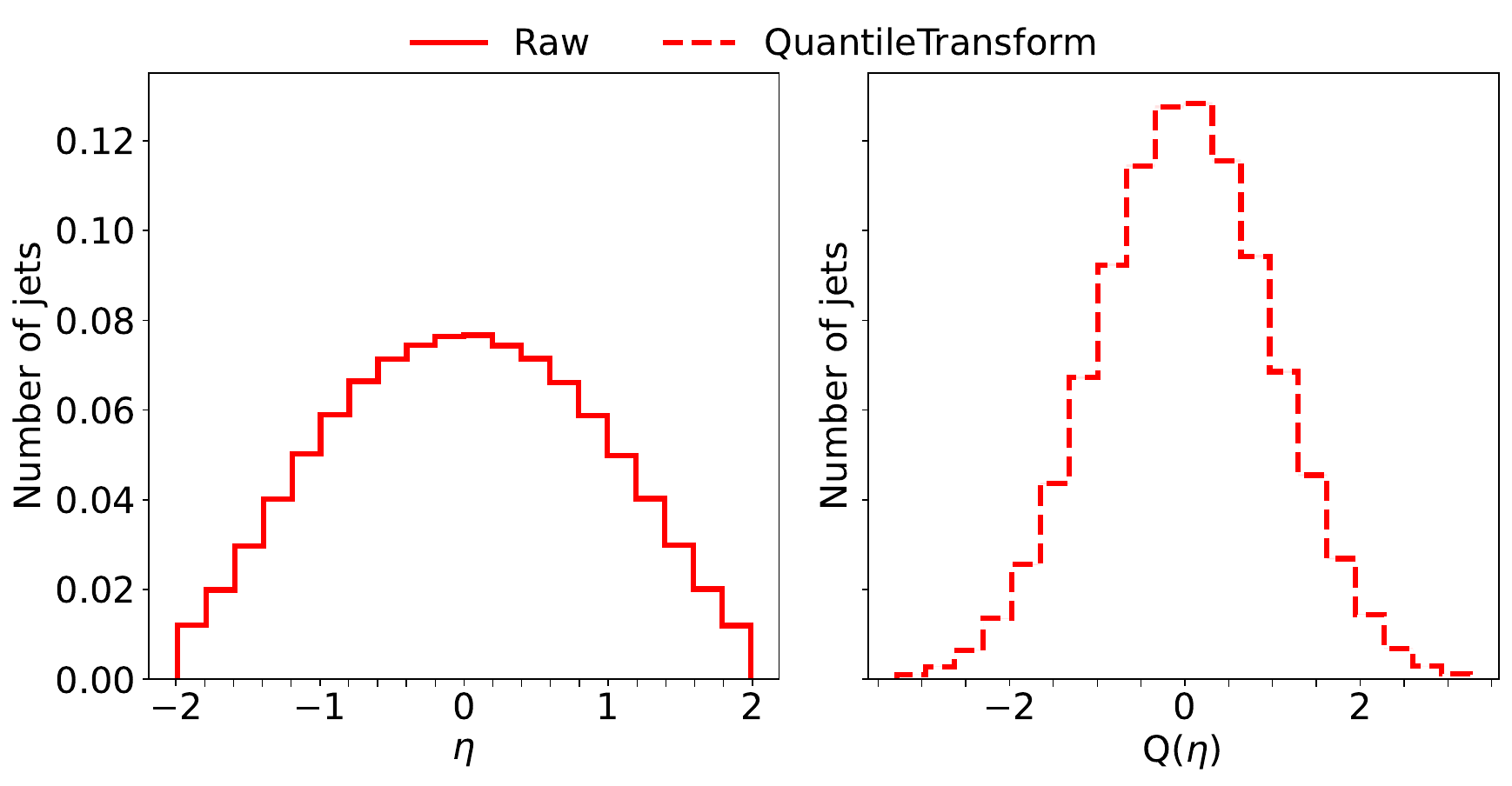}}}
    \quad
    \subfloat[]{{\includegraphics[width=0.42\textwidth]{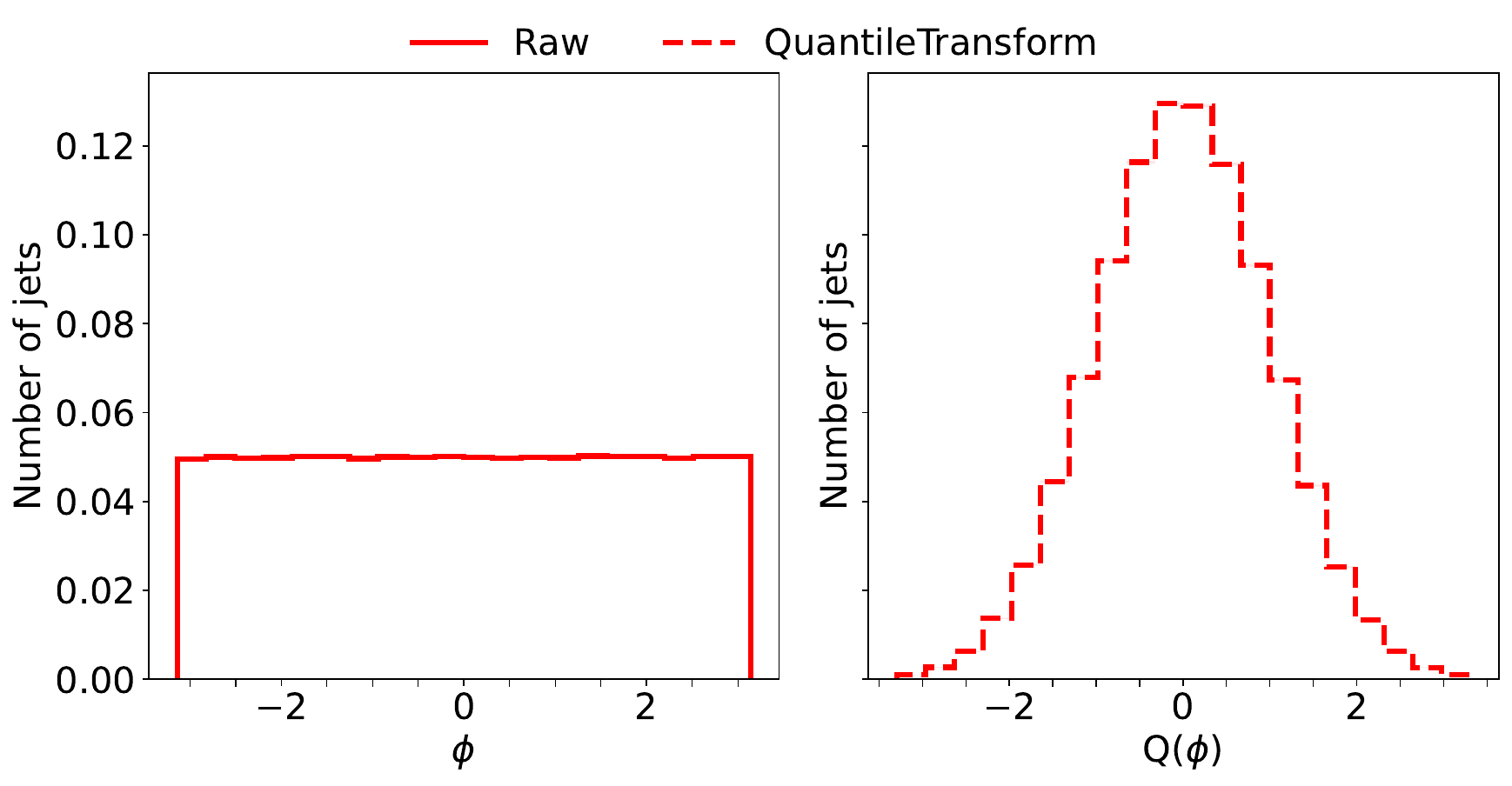}}}
    \quad
    \subfloat[]{{\includegraphics[width=0.42\textwidth]{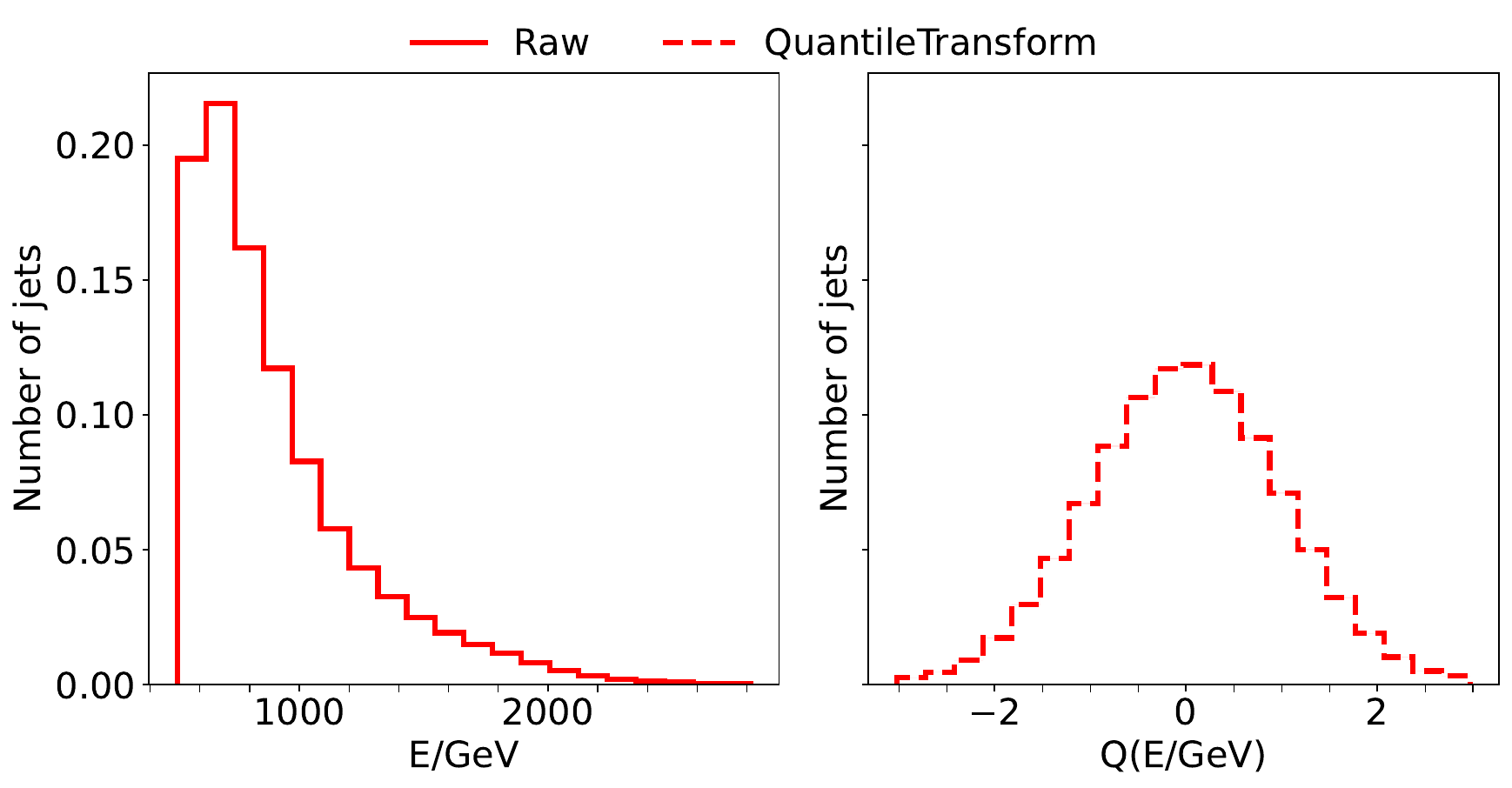}}}
    \quad
    \subfloat[]{{\includegraphics[width=0.42\textwidth]{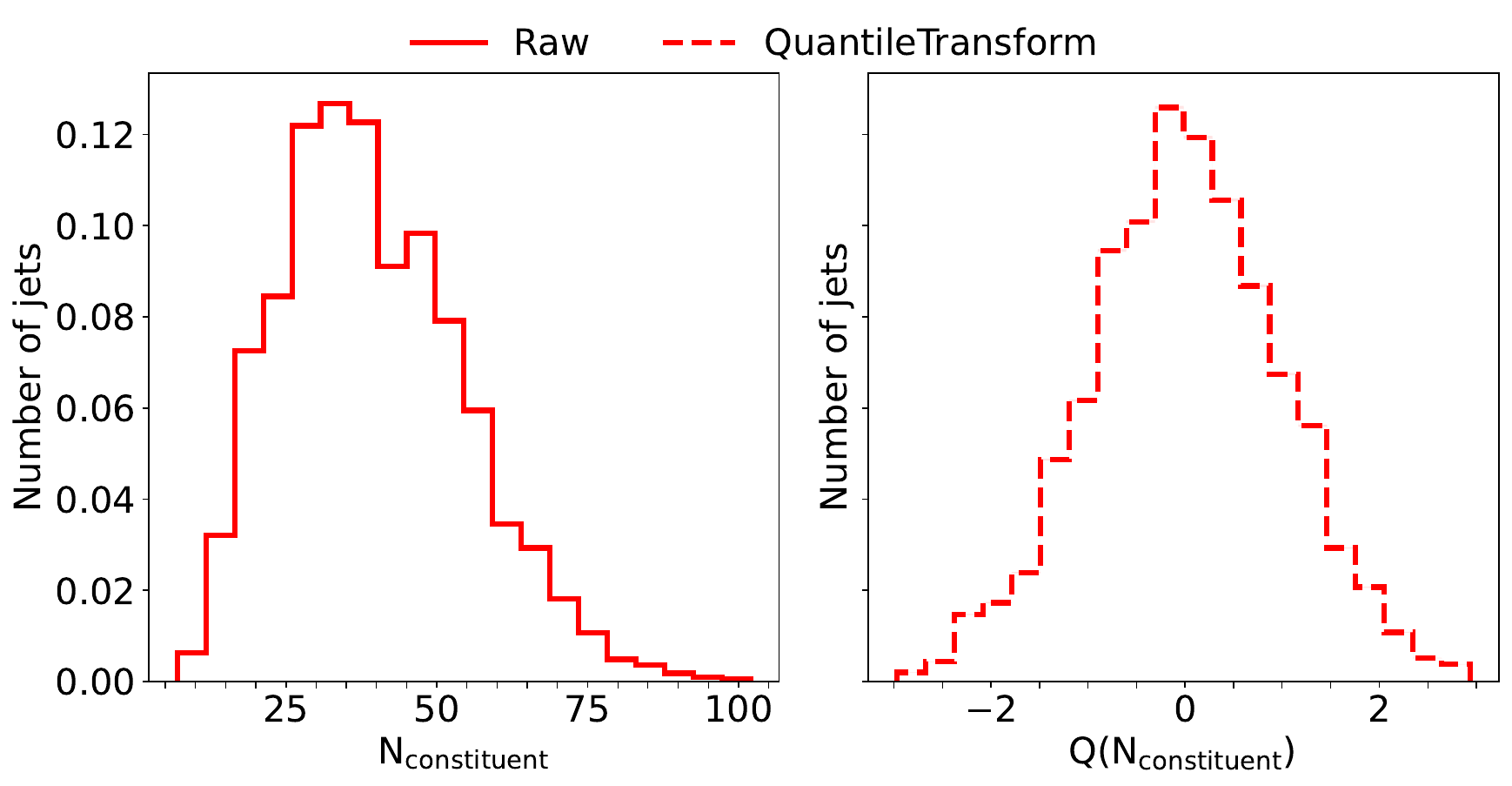}}}
    \caption{Marginal distributions of the JetClass jet features used as input to the classifier. 
    Each figure shows the marginal distribution of a single jet feature 
    and the quantile-transformed feature. 
    The quantile-transformed feature is used as input to the classifier.}
    \label{fig:classifier_jet_features}
\end{figure*}

\begin{figure*}[htpb]
    \centering
    \subfloat[]{{\includegraphics[width=0.42\textwidth]{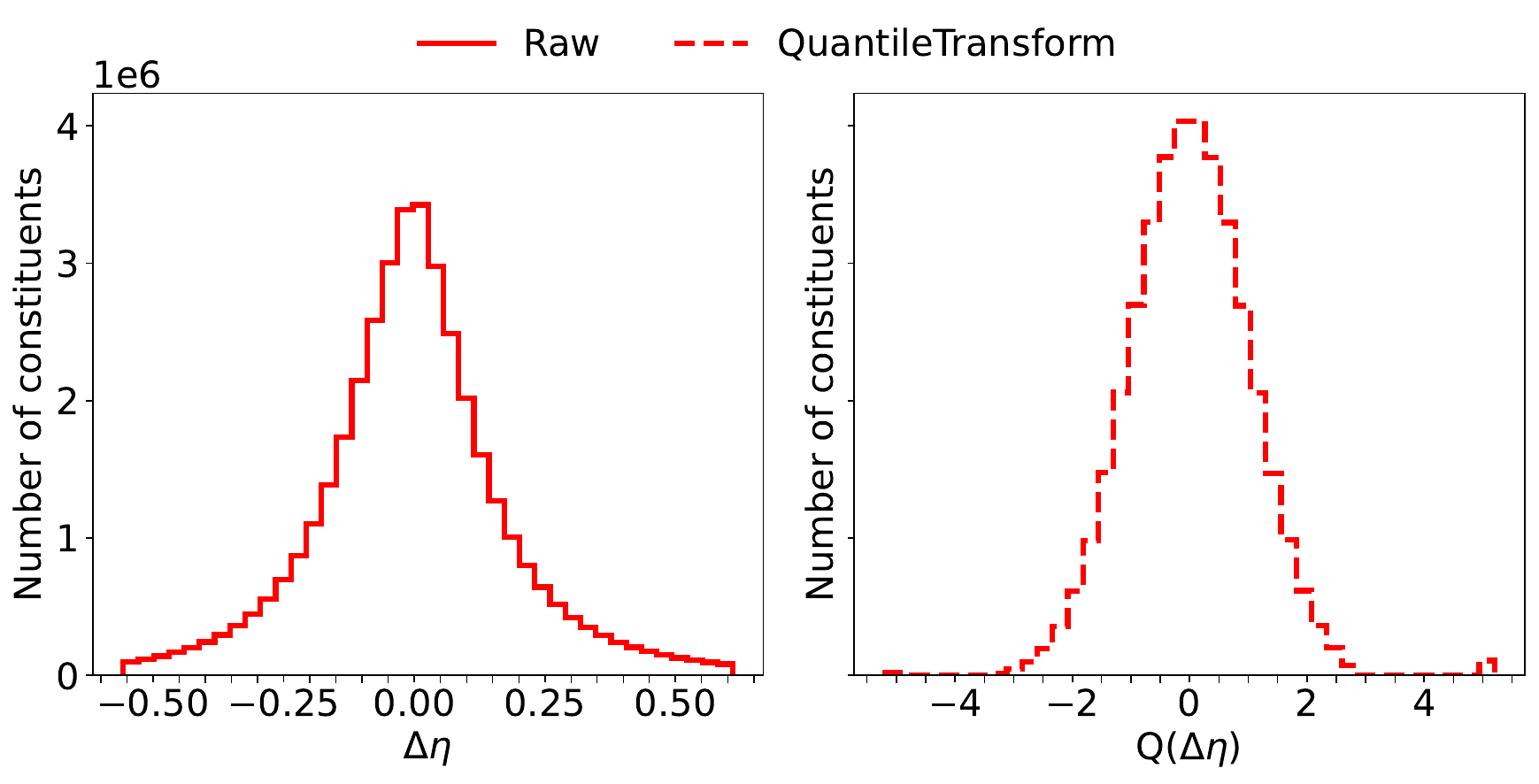}}}
    \quad
    \subfloat[]{{\includegraphics[width=0.42\textwidth]{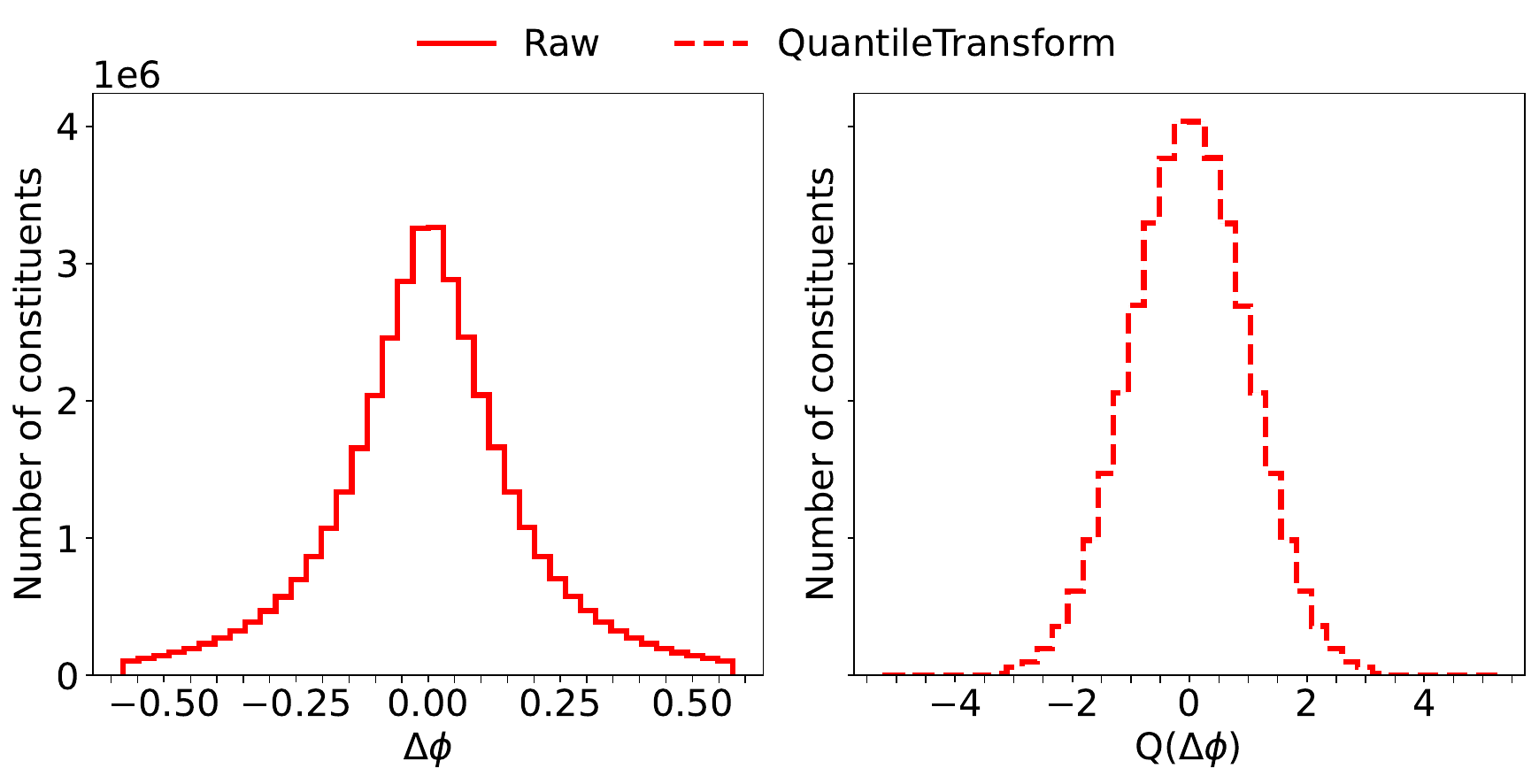}}}
    \quad
    \subfloat[]{{\includegraphics[width=0.42\textwidth]{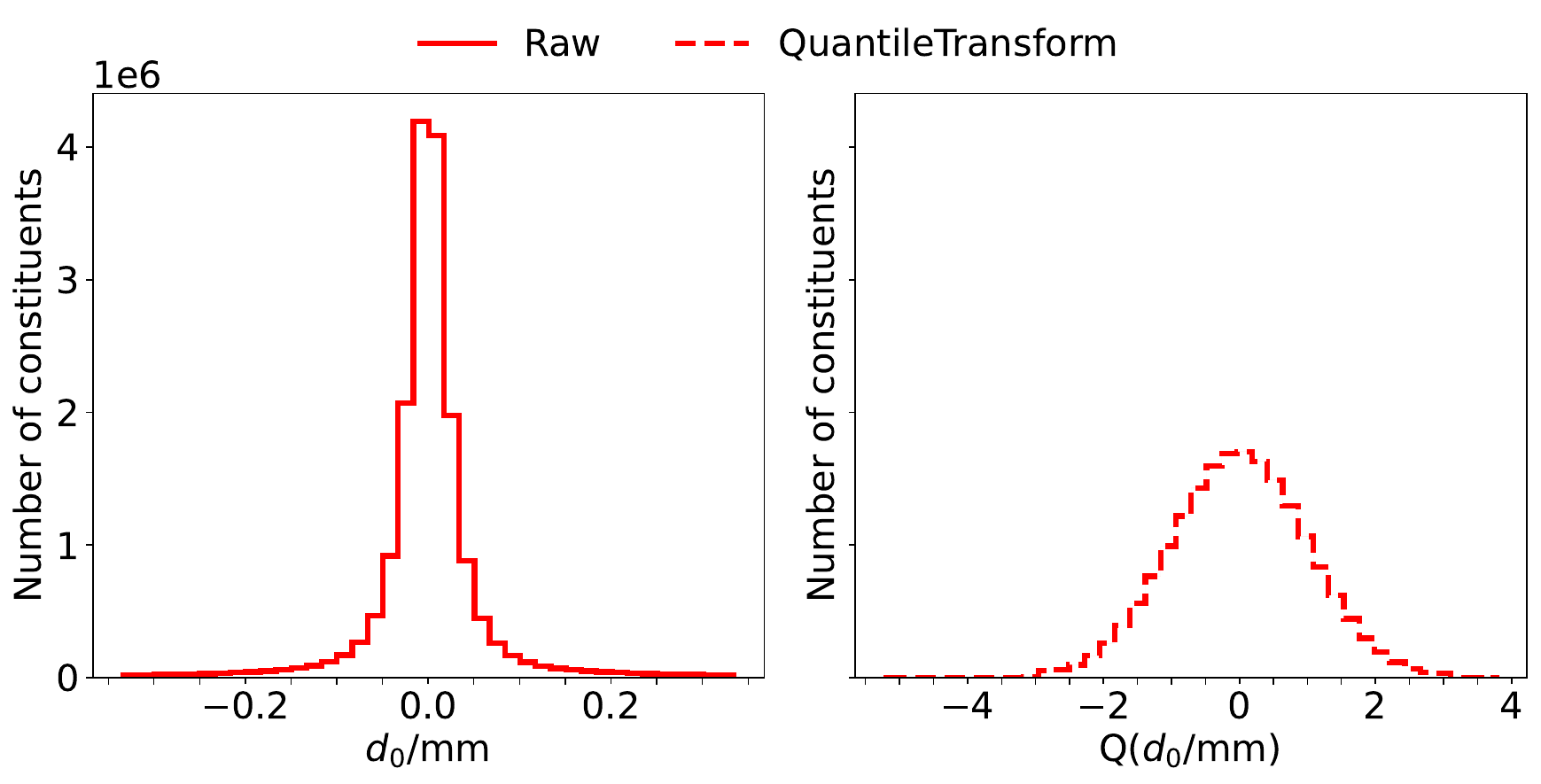}}}
    \quad
    \subfloat[]{{\includegraphics[width=0.42\textwidth]{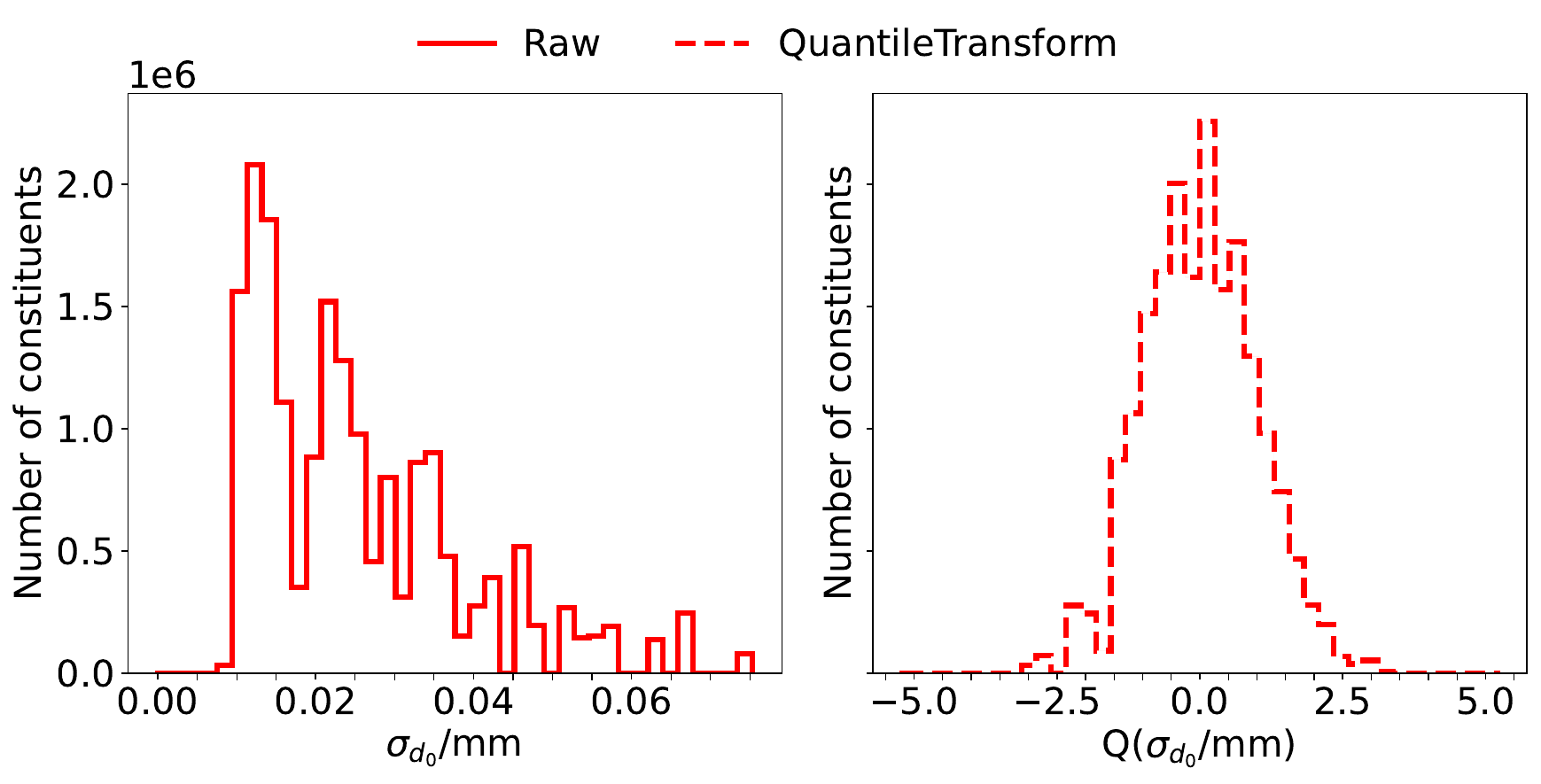}}}
    \quad
    \subfloat[]{{\includegraphics[width=0.42\textwidth]{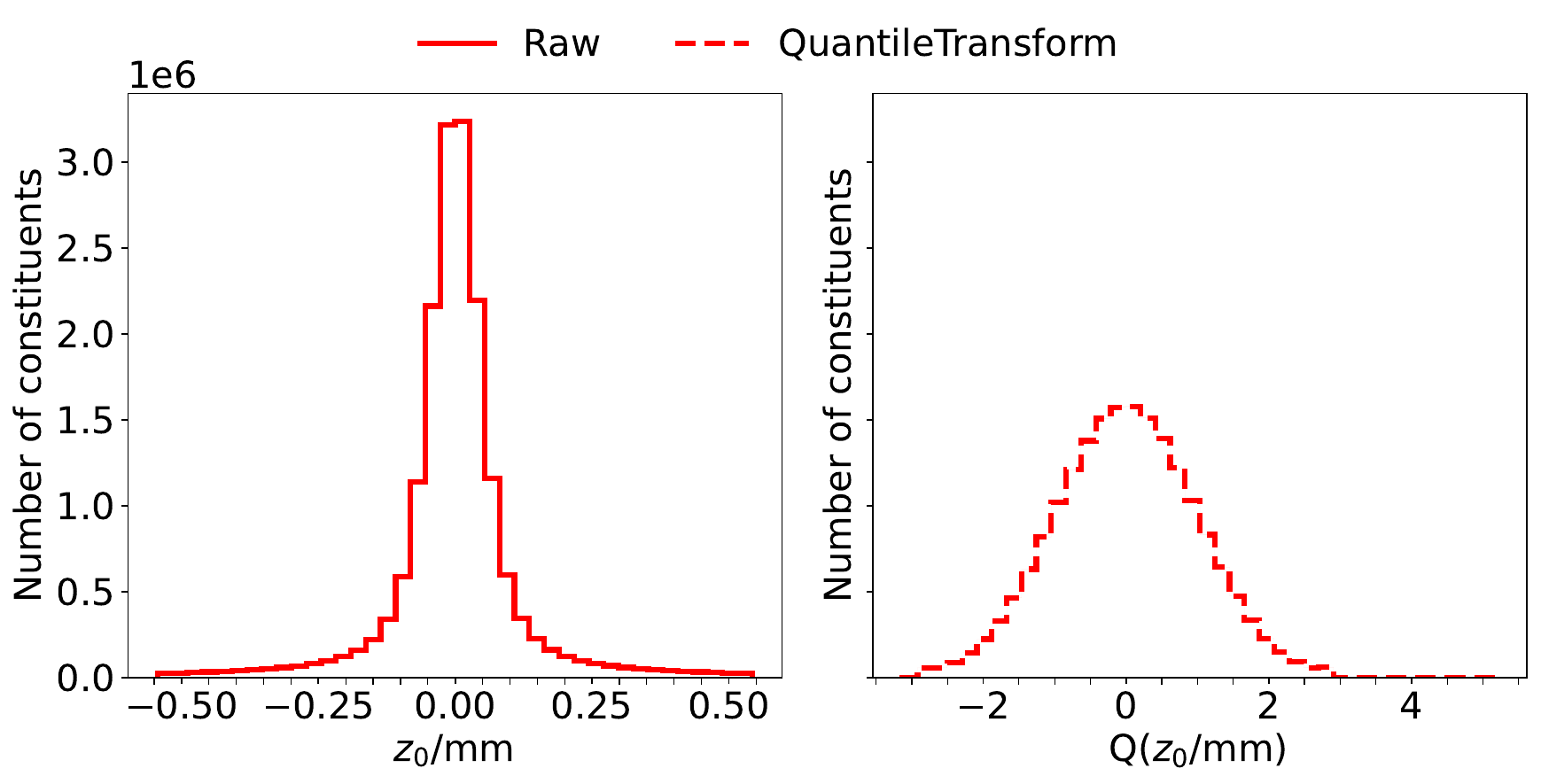}}}
    \quad
    \subfloat[]{{\includegraphics[width=0.42\textwidth]{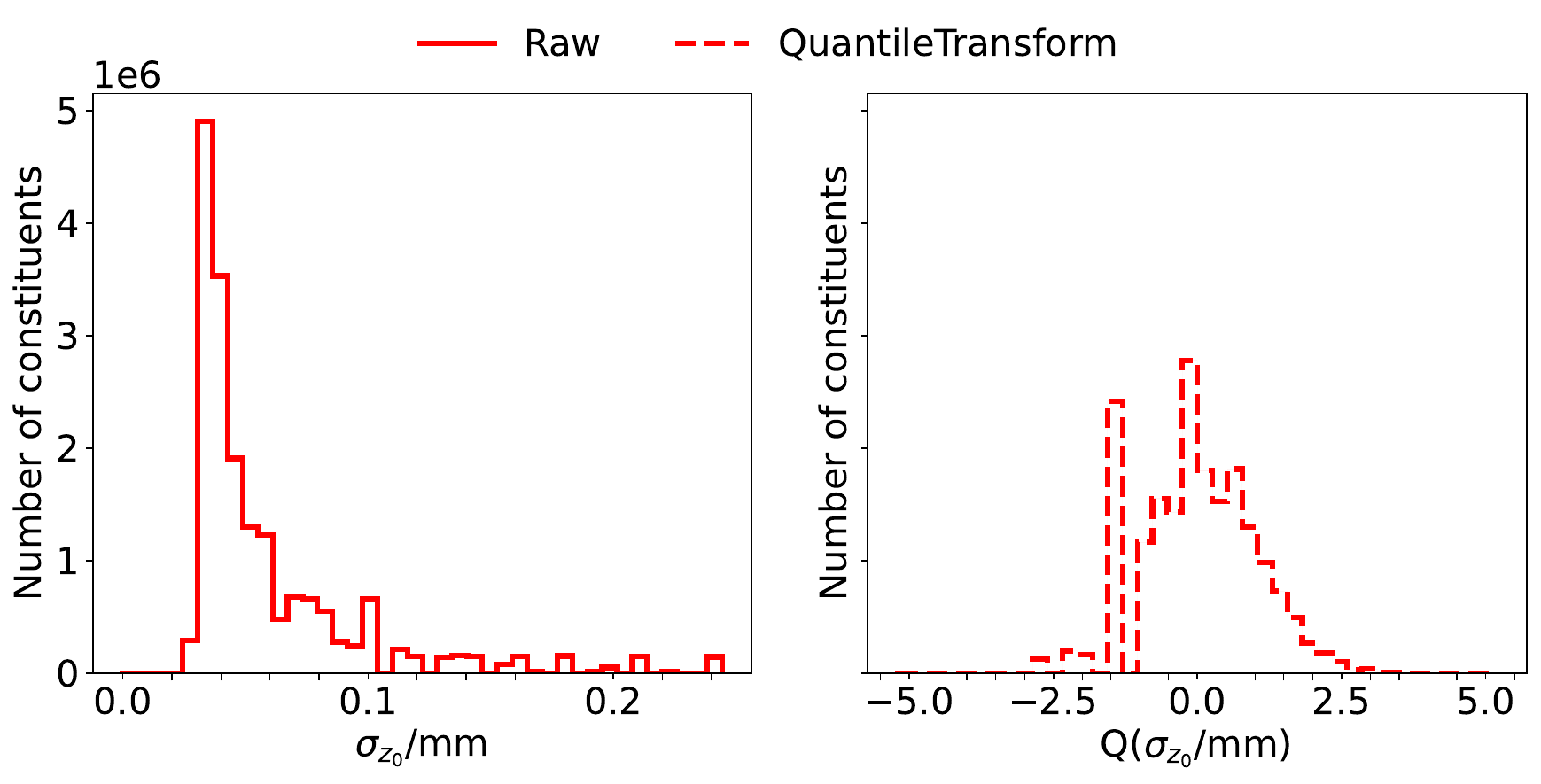}}}
    \quad
    \subfloat[]{{\includegraphics[width=0.42\textwidth]{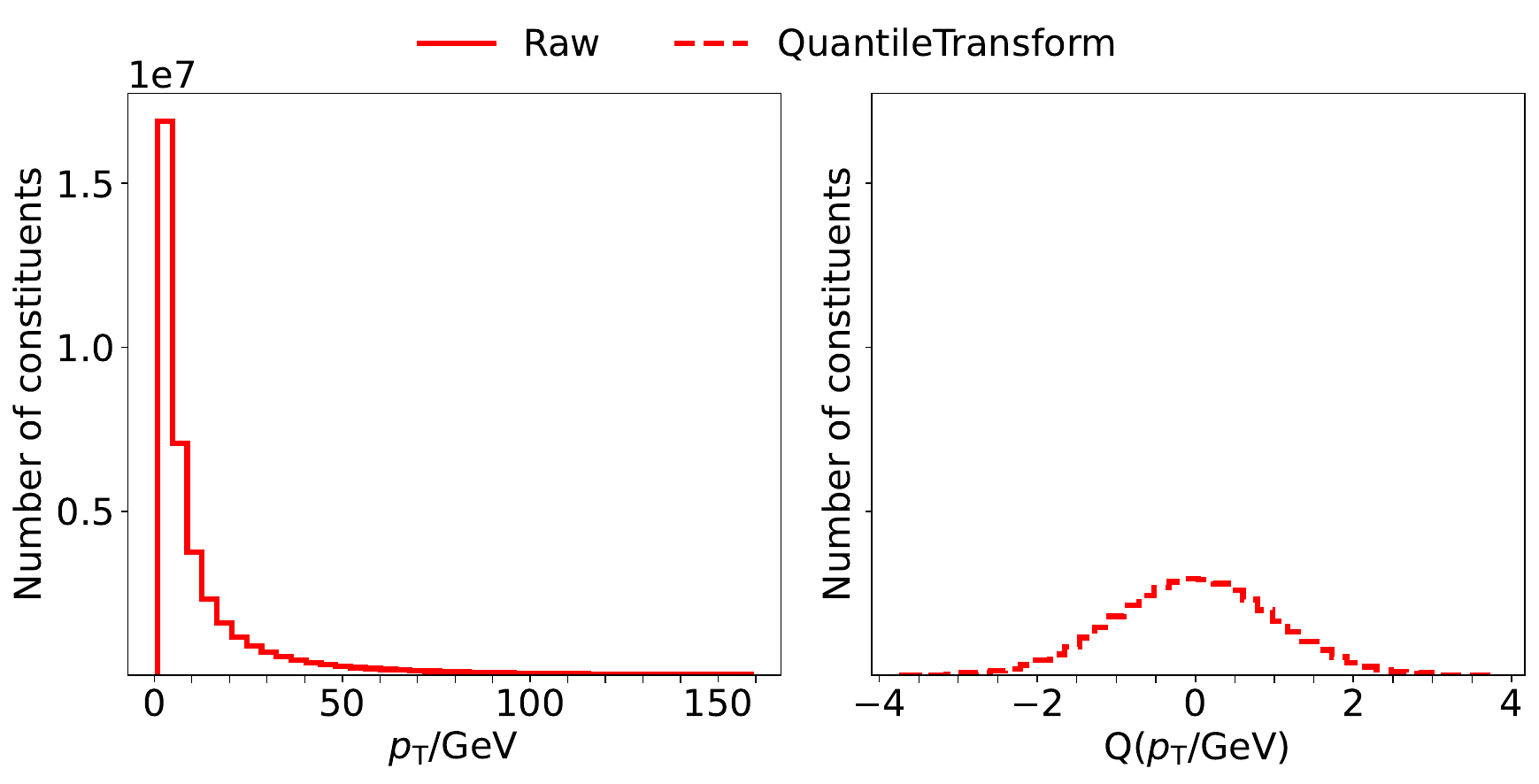}}}
    \caption{Marginal distributions of the JetClass constituent features used as input to the classifier. 
    Each figure shows the marginal distribution of a single constituent feature 
    and the quantile-transformed feature. 
    The quantile-transformed feature is used as input to the classifier.
    For some features, they are not defined for neutral particles. In those features the neutrals are removed.}
    \label{fig:classifier_csts_features}
\end{figure*}

\cref{fig:classifier_jet_features,fig:classifier_csts_features} show the marginal distributions of the input features normalized with $QuantileTransformer$.

\section{Results}

\subsection{Calibrated results for \OMLP} \label{sec:ref2_comment}
\begin{figure*}[htpb]
    \centering
    \subfloat[]{{\includegraphics[width=0.45\textwidth]{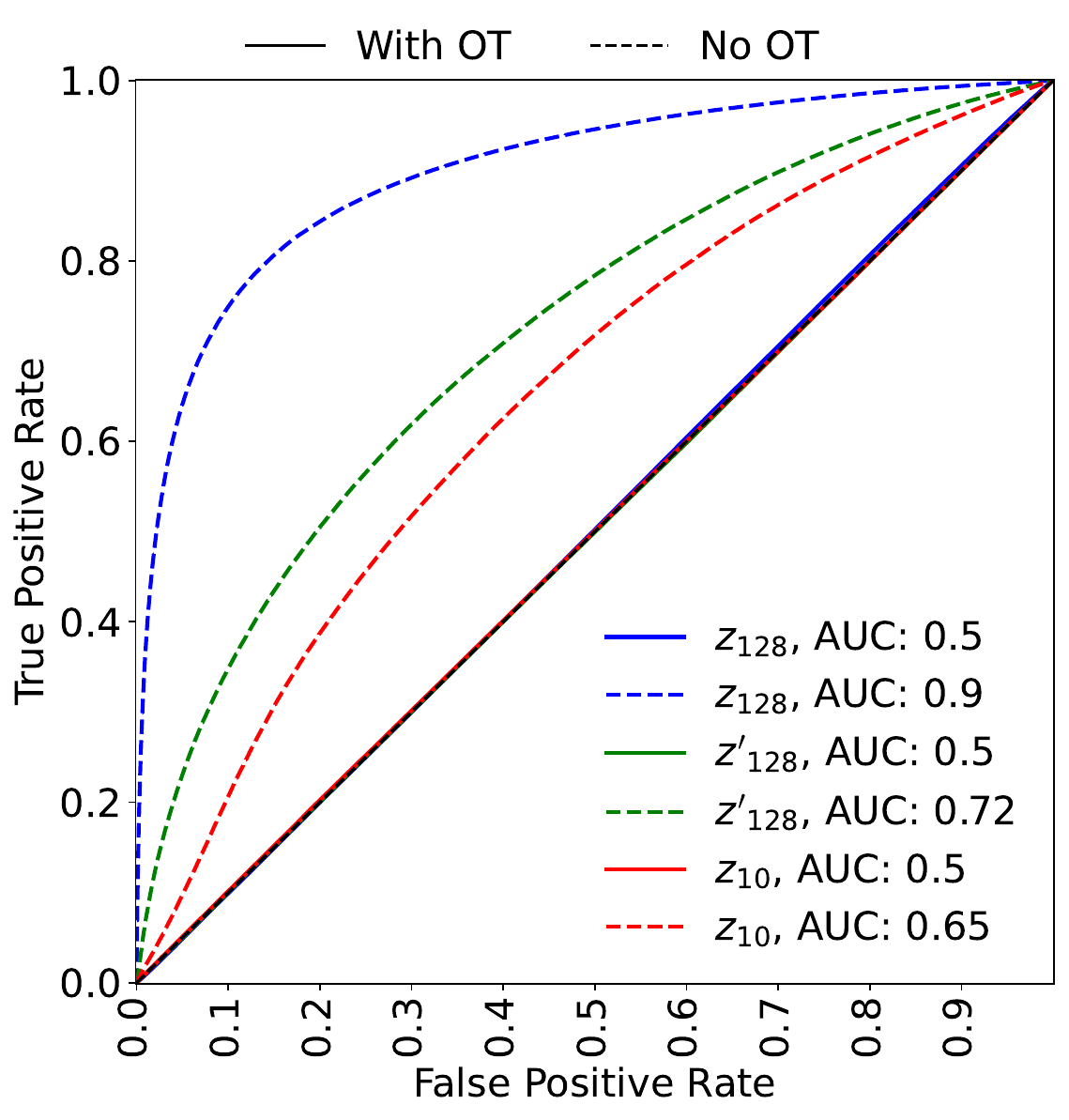}}}
    \subfloat[]{{\includegraphics[width=0.45\textwidth]{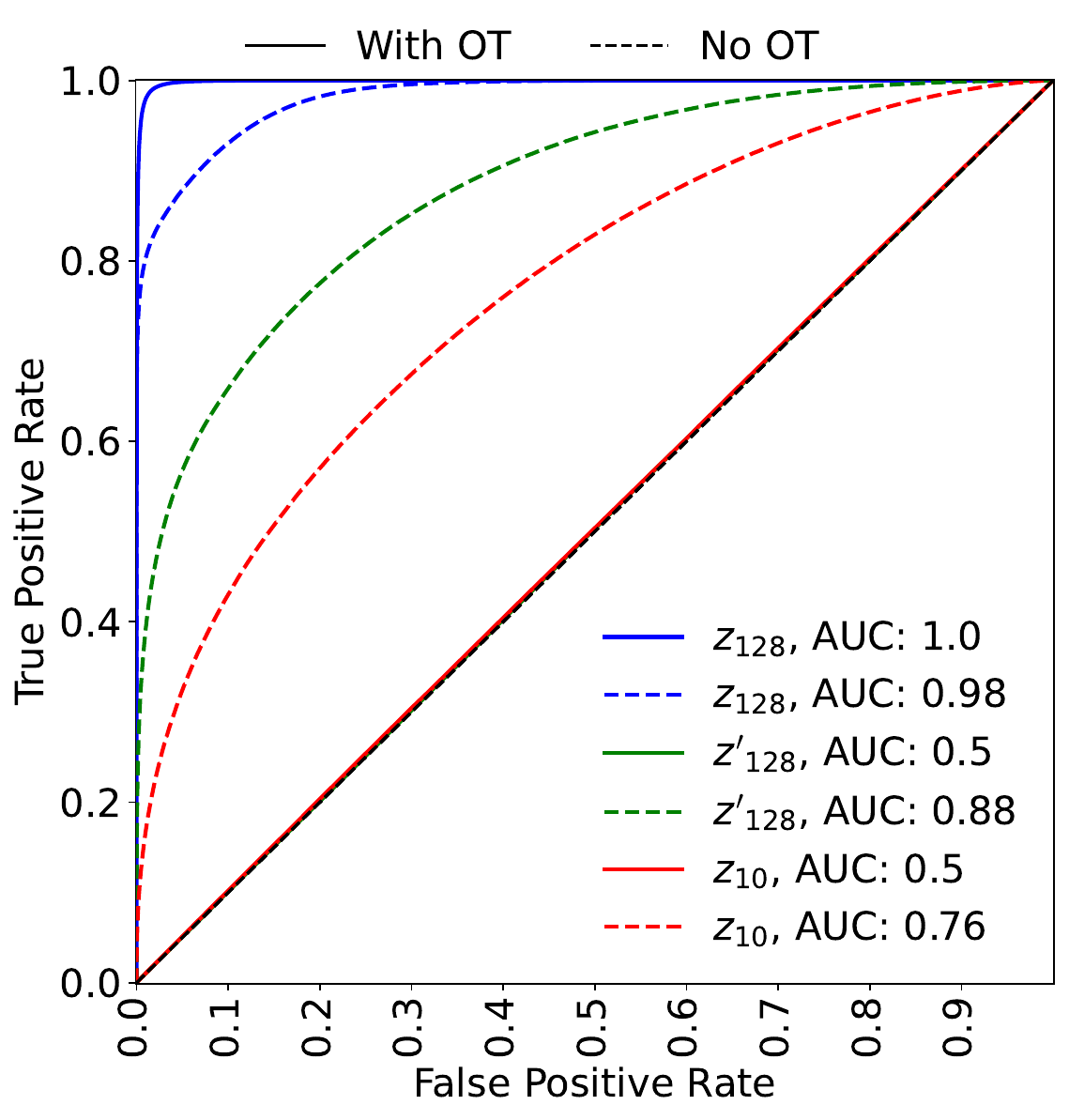}}}
    \caption{ROC curves of discriminators trained on the $z_{128}$, $z_{128}^{'}$ and $z_{10}^{'}$ latent space. Here the calibration has been derived in $z_{128}$ and the later layers in the \OMLP have been applied to map it to $z_{128}^{'}$ and $z_{10}^{'}$. 
    Given the linestyle of the ROC curves, 
    the discriminators are either trained to distinguish between the calibration and the target or nominal vs target. 
    Each AUC is also included in the legend.
    (a) shows the ROC curves of a single linear layer classifier. 
    (b) shows the ROC curves of \DL.
}
    \label{fig:only_128d_rocs_appendix}
\end{figure*}

\begin{figure*}[htpb]
    \centering
    \subfloat[]{{\includegraphics[width=0.30\textwidth]{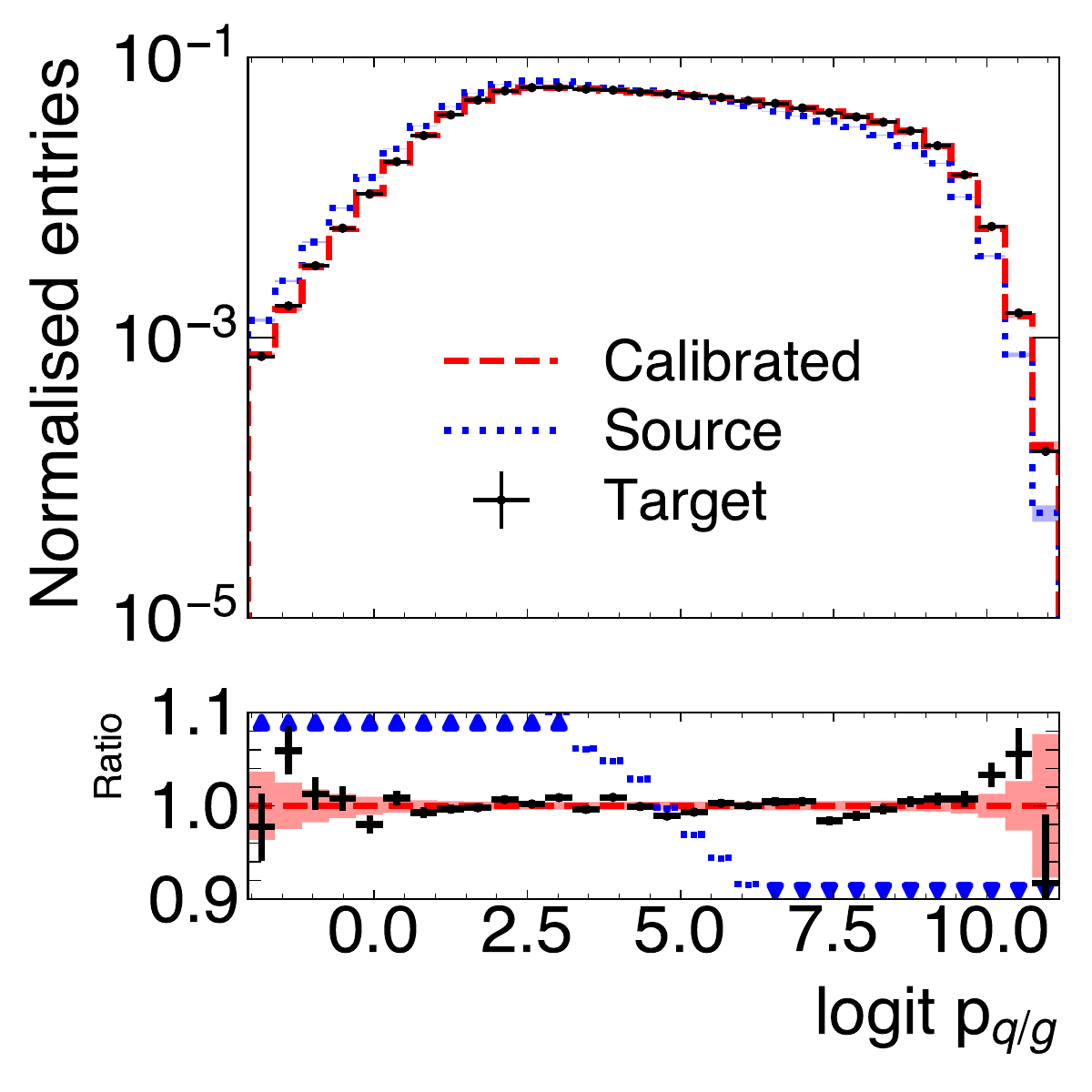}}}
    \subfloat[]{{\includegraphics[width=0.30\textwidth]{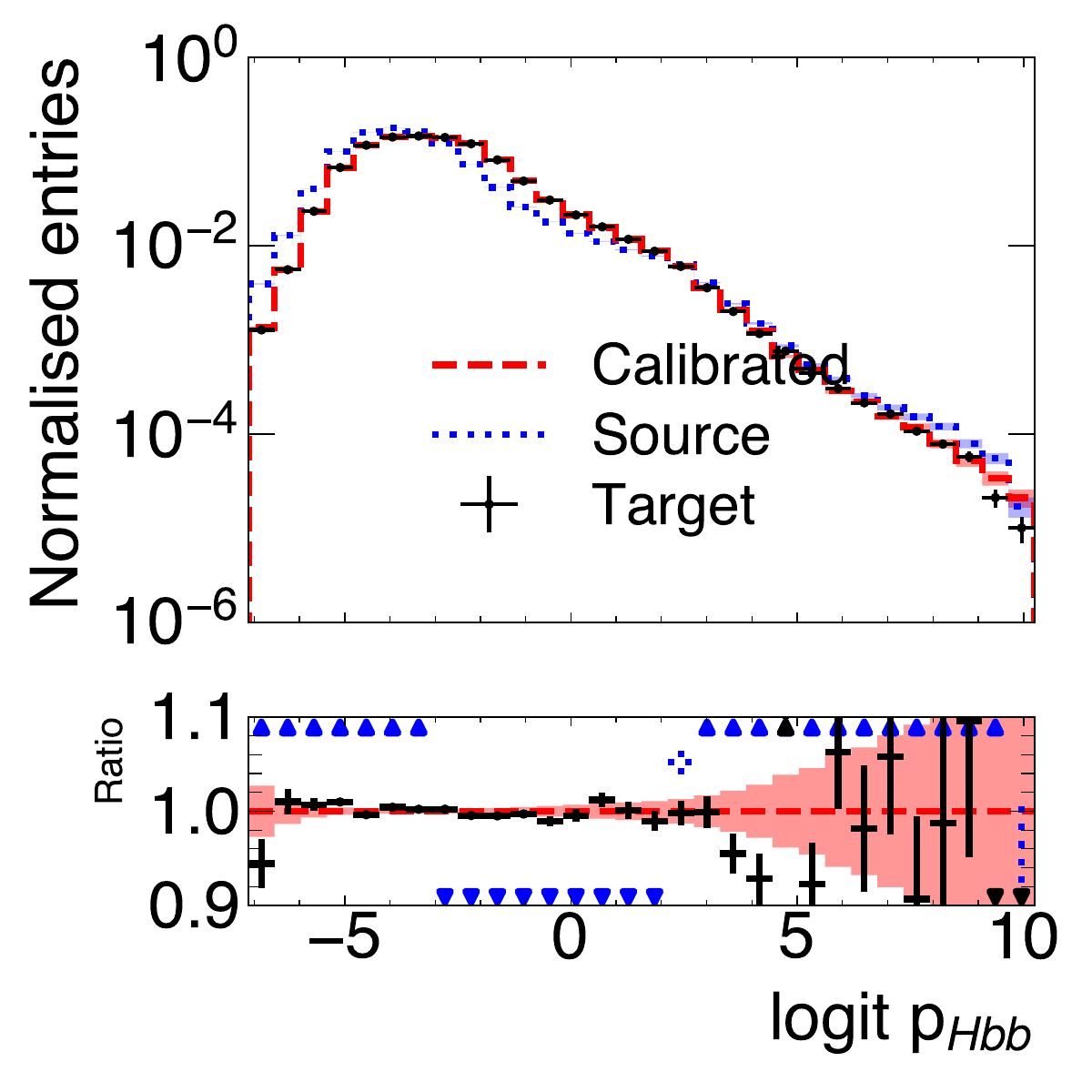}}}
    \subfloat[]{{\includegraphics[width=0.30\textwidth]{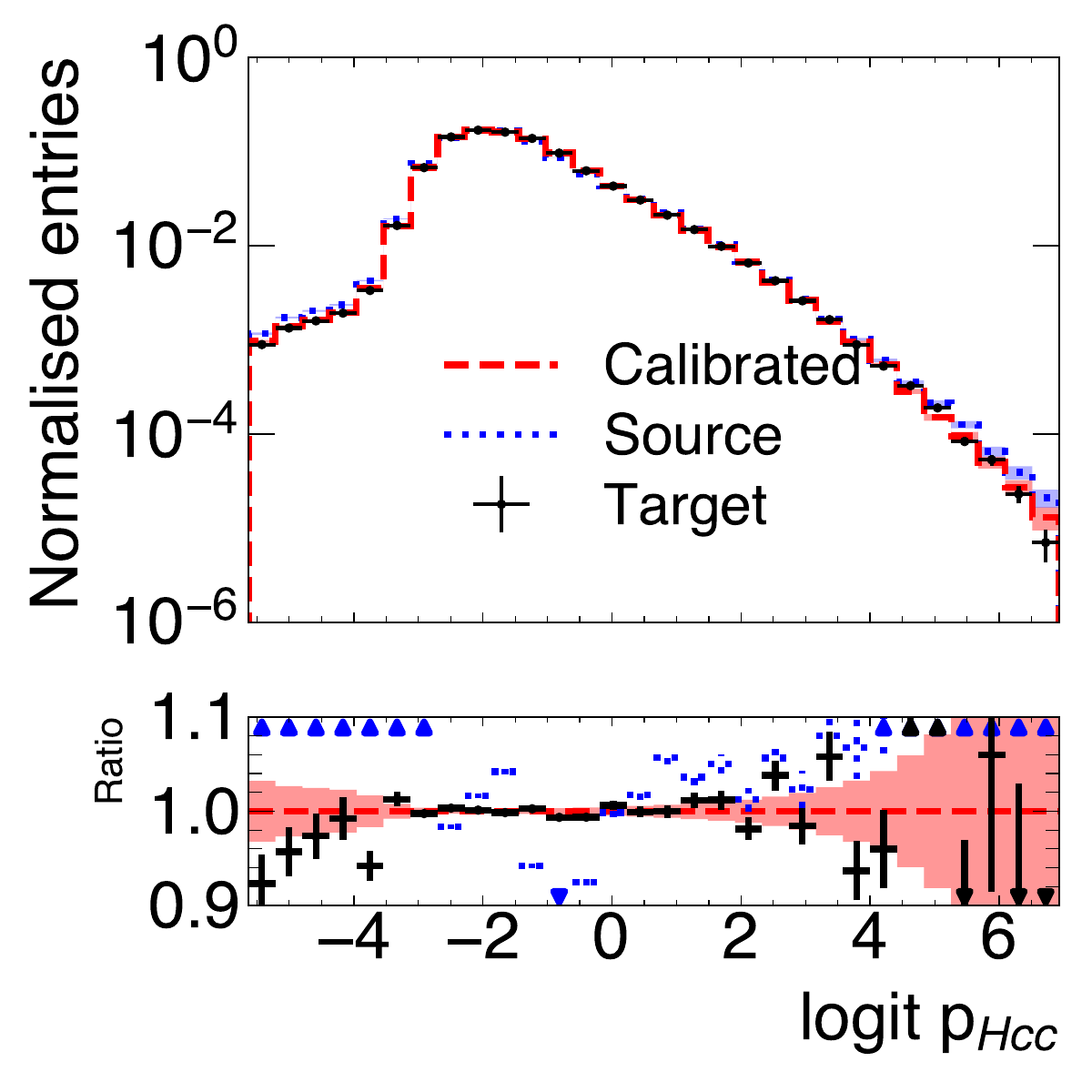}}}
    \\
    \subfloat[]{{\includegraphics[width=0.30\textwidth]{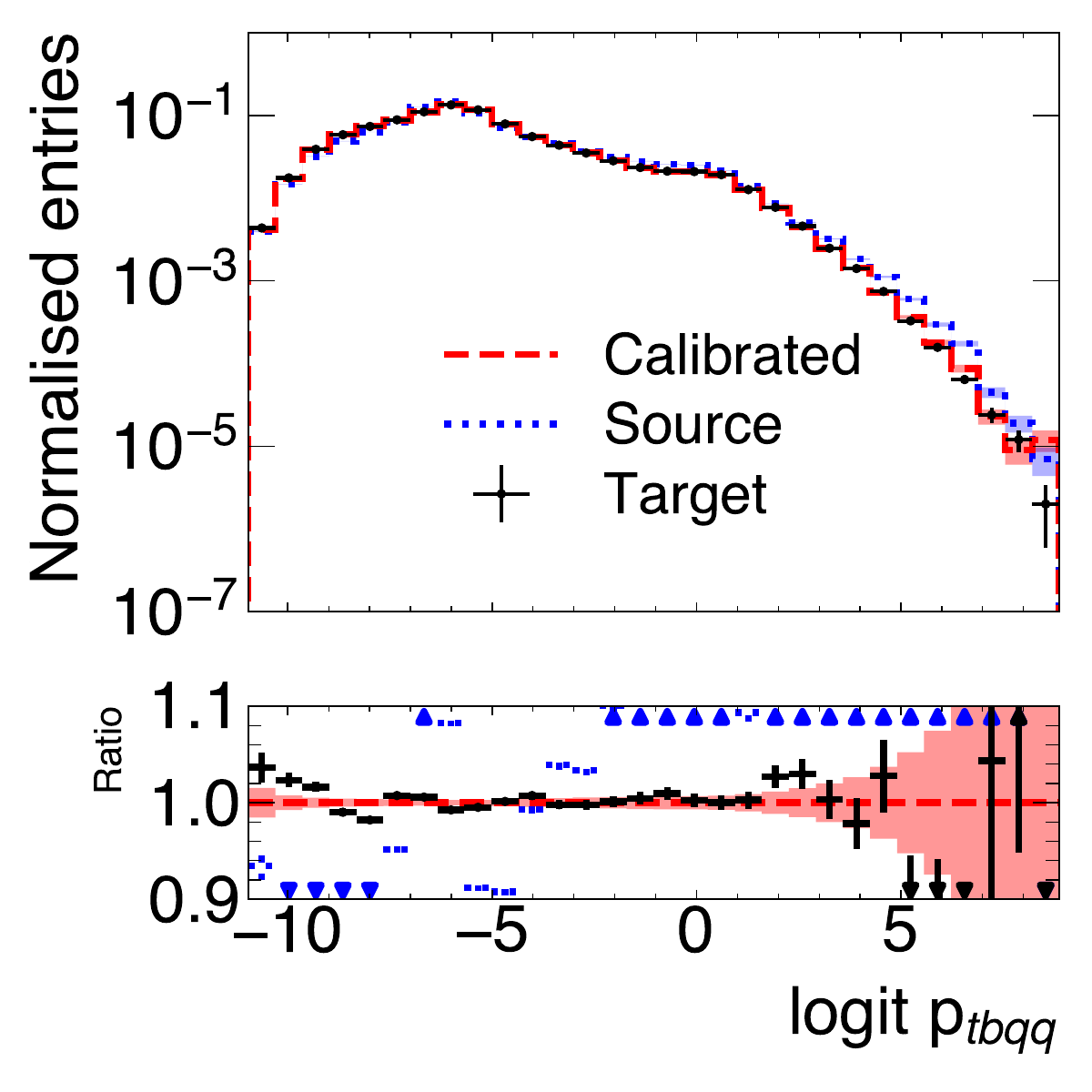}}}
    \subfloat[]{{\includegraphics[width=0.30\textwidth]{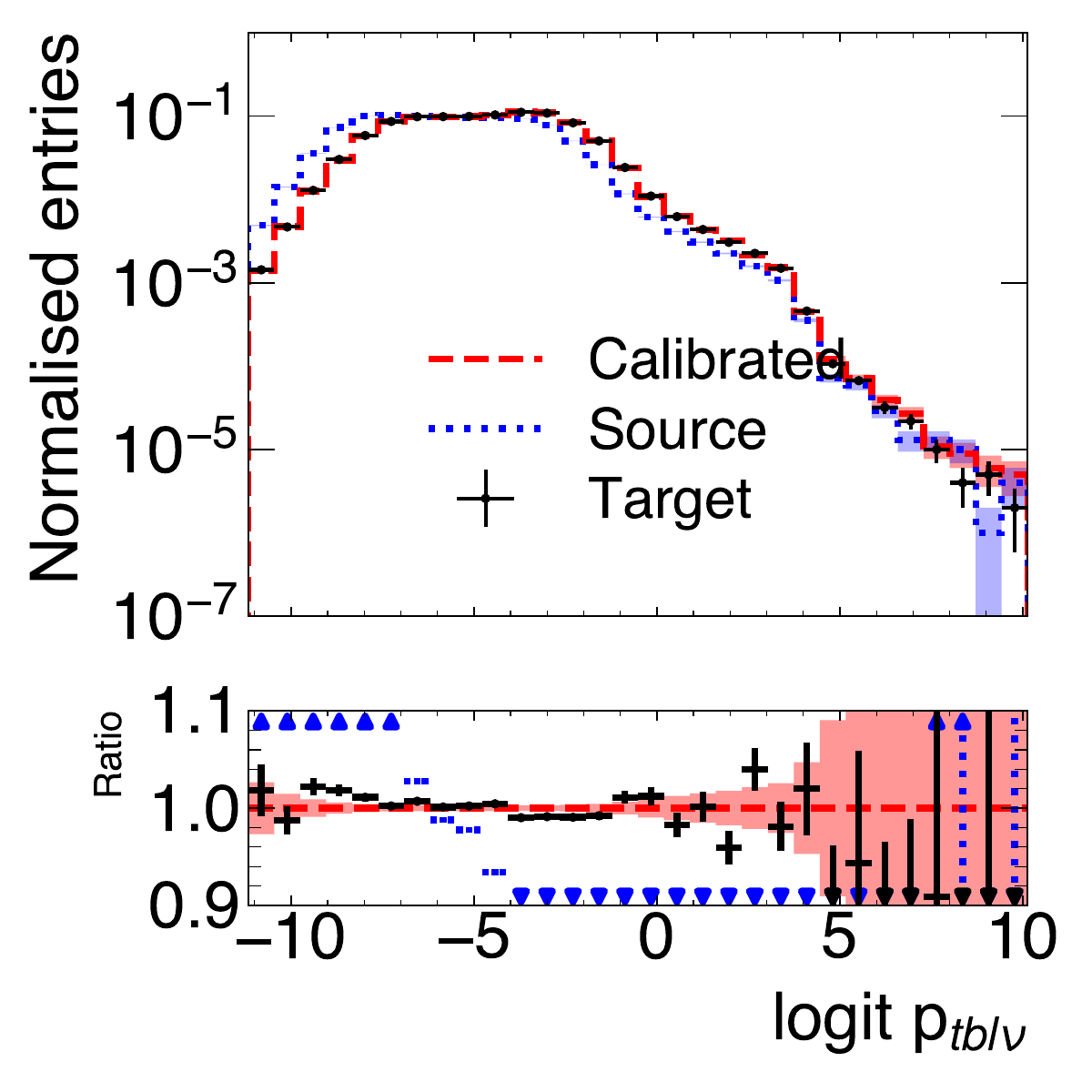}}}
    \subfloat[]{{\includegraphics[width=0.30\textwidth]{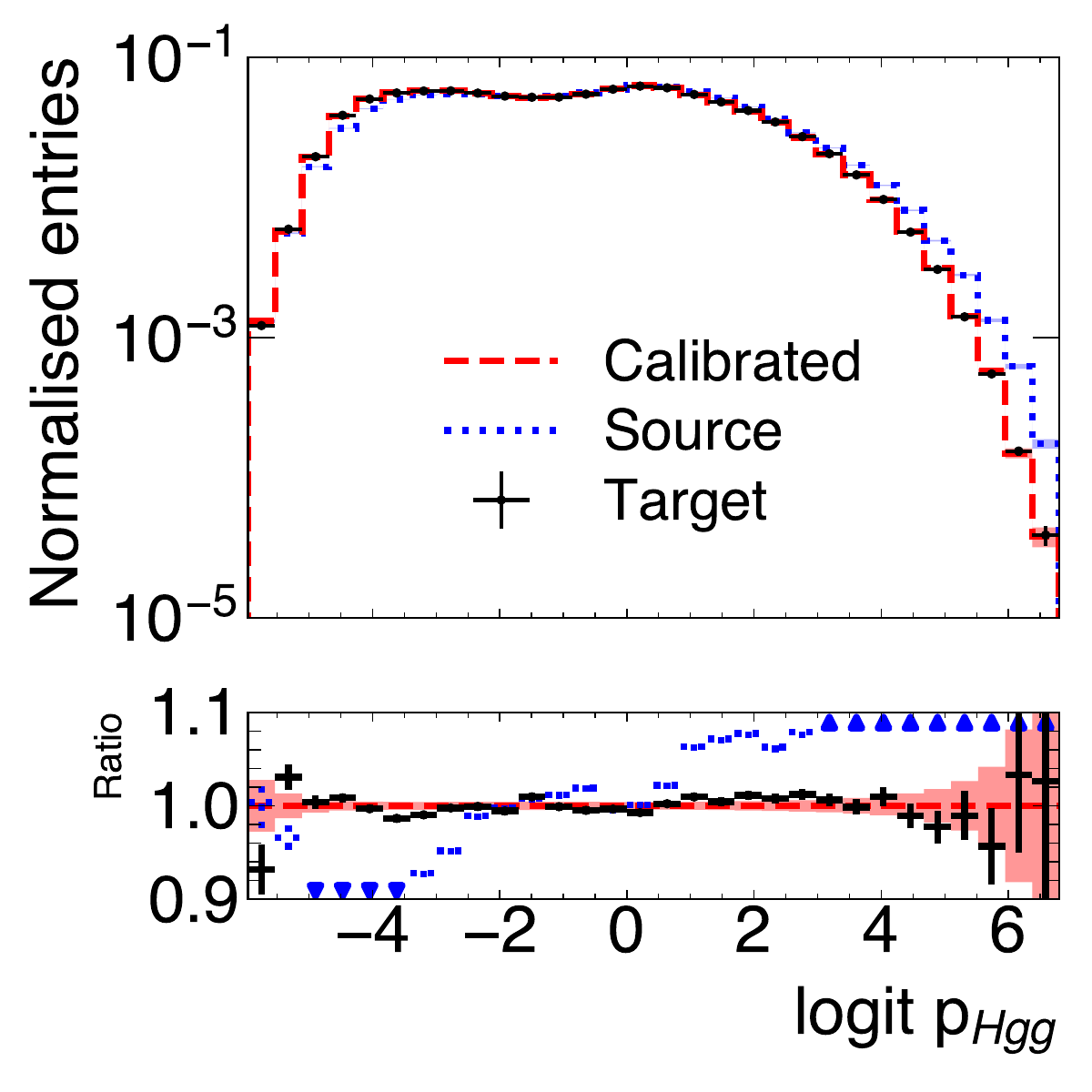}}}
    \\
    \subfloat[]{{\includegraphics[width=0.30\textwidth]{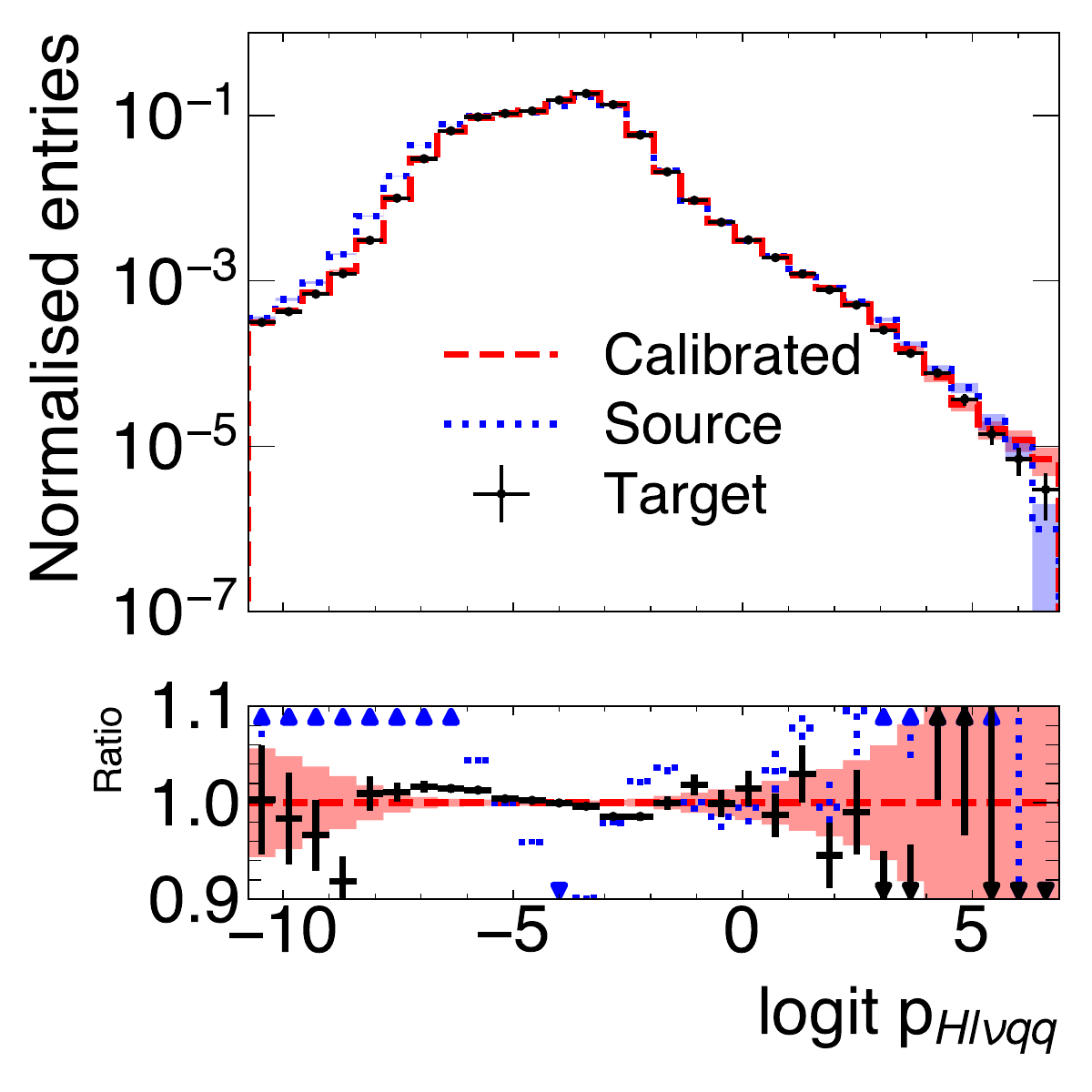}}}
    \subfloat[]{{\includegraphics[width=0.30\textwidth]{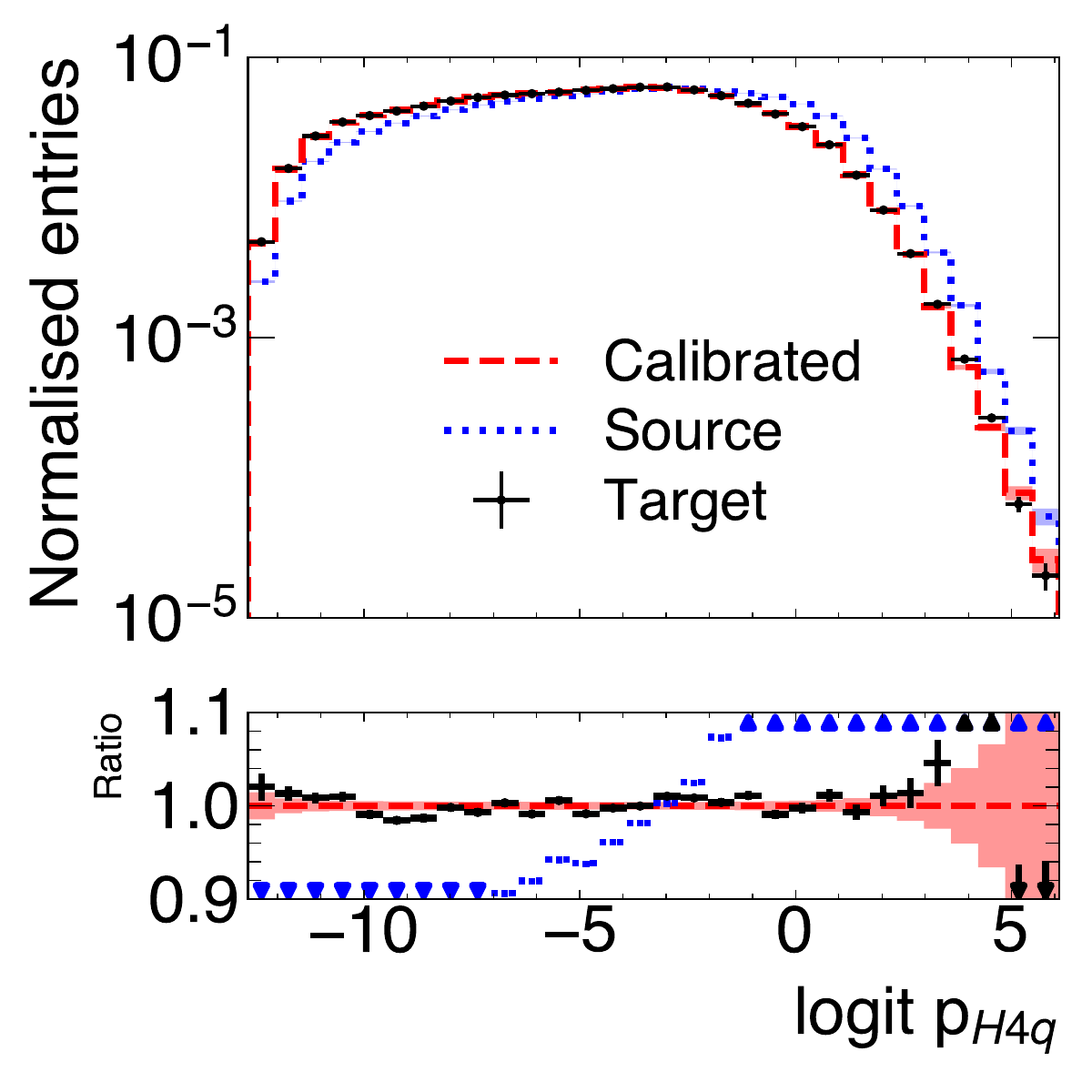}}}
    \subfloat[]{{\includegraphics[width=0.30\textwidth]{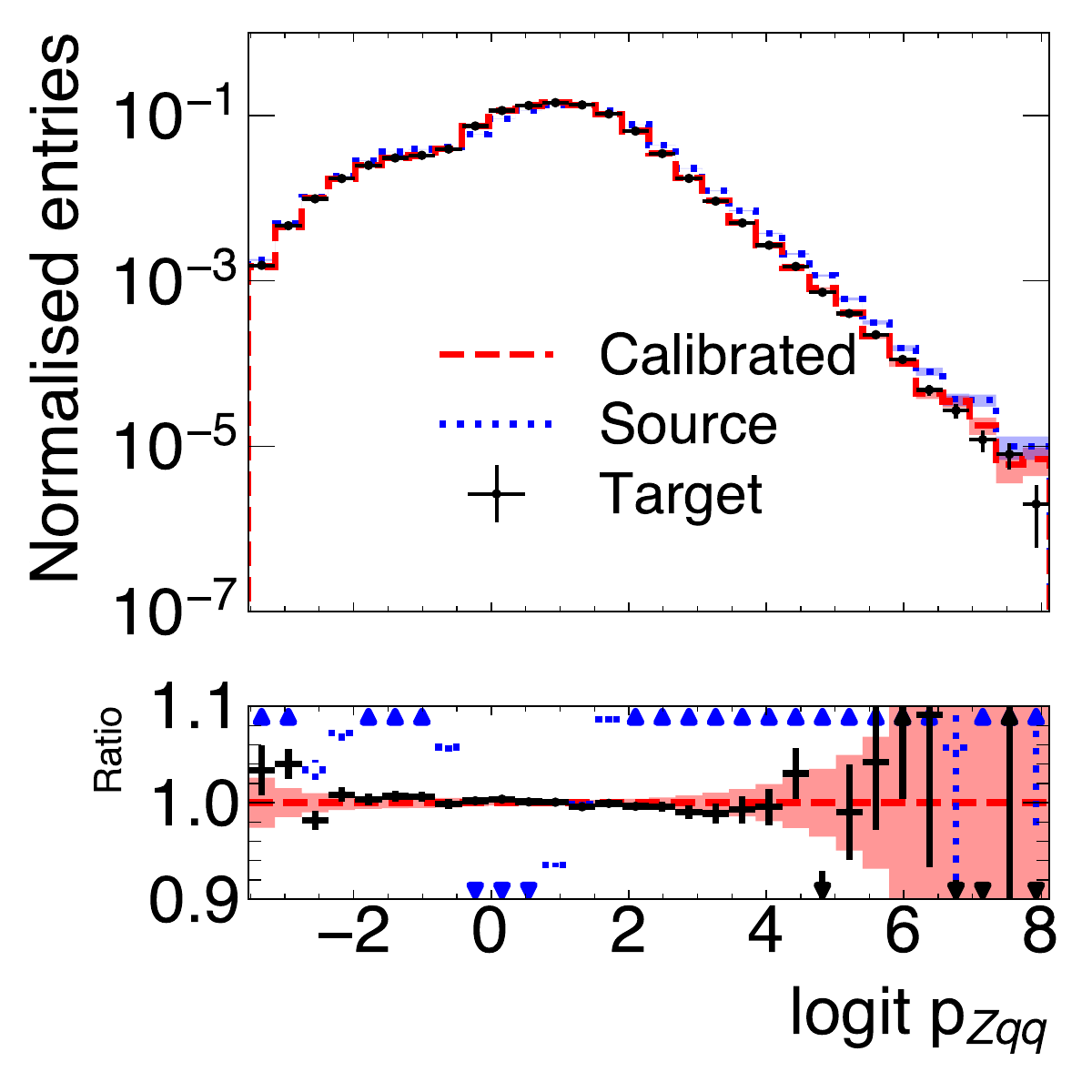}}}
    \\
    \subfloat[]{{\includegraphics[width=0.30\textwidth]{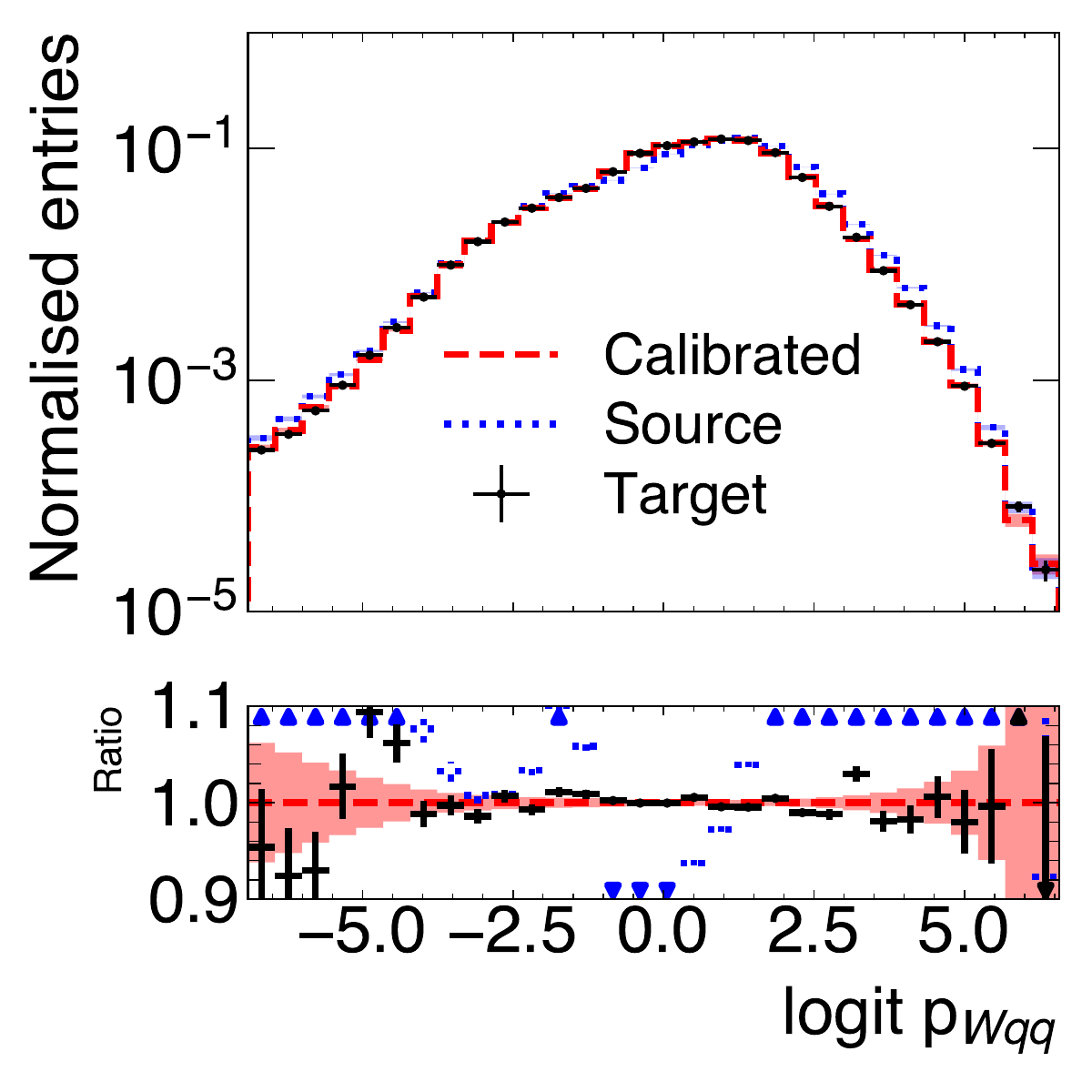}}}
    \caption{
    Comparisons of the marginal distributions of the output space $z_{10}^{'}$ of the classifier.
    The black distribution represents target distribution, red is the transported distribution and blue is the source distribution.
    The transport has been derived in $z_{128}$ and the later layers in the \OMLP have been applied to map it to $z_{10}^{'}$. 
    Closure of the calibration is indicated by the difference to the black distribution.
    Description of the labels can be found in \cref{eq:labels}.
    }
    \label{fig:only_128d_marginal_distribution_output_space_appendix}
\end{figure*}

\begin{figure*}[htpb]
    \centering
    \includegraphics[width=0.50\textwidth]{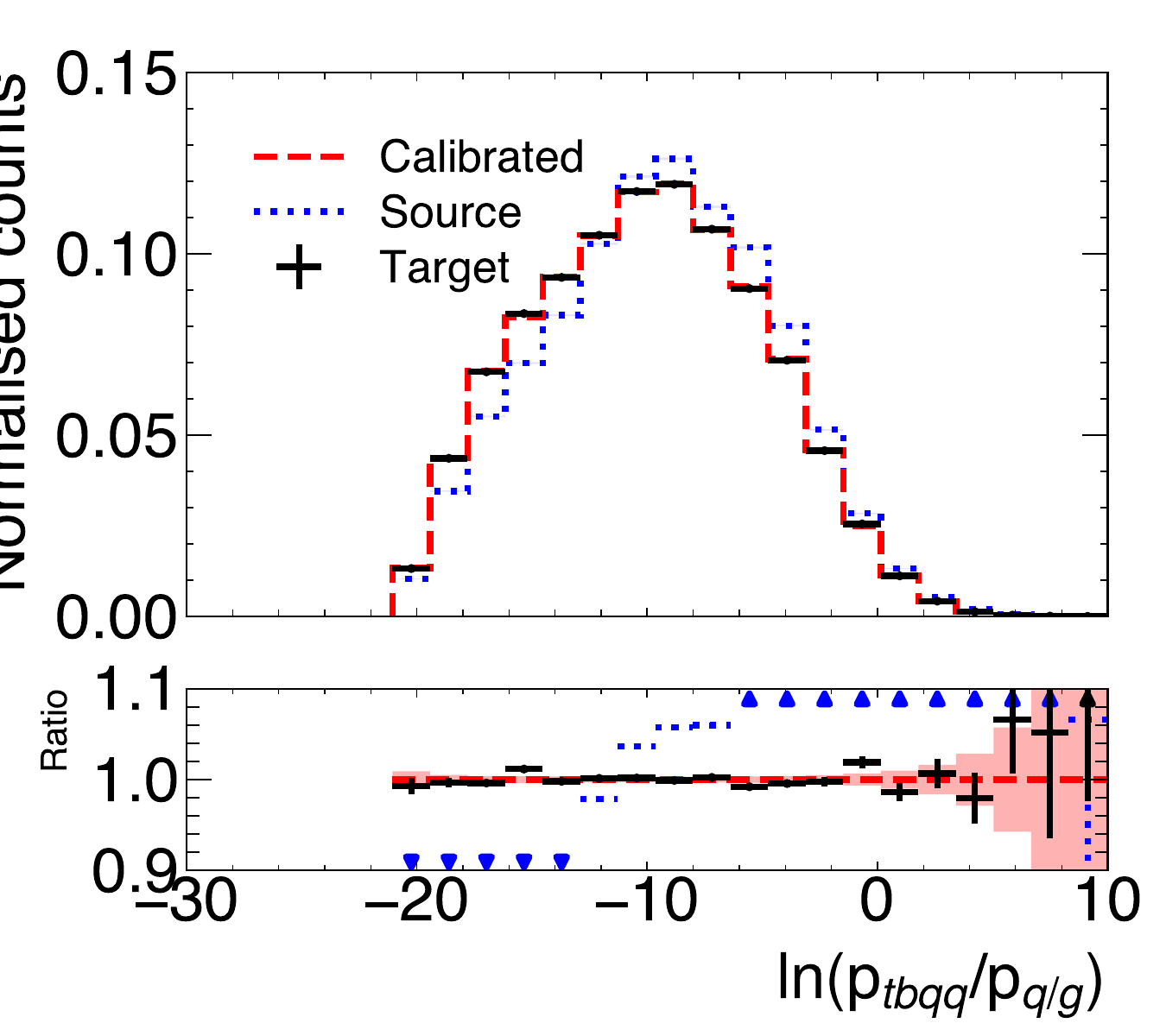}
    \caption{
        Comparison between the marginal distributions of a physics-motivated
        one-dimensional discriminant between hadronic top-quark decays and
        QCD multijet.
        The distributions are represented as follows: target distribution
        (black), calibrated distribution (red), and source distribution (blue). 
        The calibration is derived on the $z_{128}$ latent space, then the
        subsequent \OMLP layers are applied to map it to output space
        $z_{10}^{'}$ where Softmax is applied to map the output space to a
        simplex. 
        Afterwards the output space is projected to physics-motivated
        one-dimensional discriminants. 
        Description of the labels can be found in \cref{eq:labels}.
    }
    \label{fig:only_128d_marginal_distribution_physics_dics_appendix}
\end{figure*}

\cref{fig:only_128d_rocs_appendix} show the ROC curves of the discriminators trained to discriminate between the calibration and target distributions, as well as the nominal and target distributions.
\cref{fig:only_128d_marginal_distribution_output_space_appendix} show the marginal distributions of the output space of the classifier, 
comparing the target distribution, the transported distribution, and the source distribution.
\cref{fig:only_128d_marginal_distribution_physics_dics_appendix} shows an additional physics motivated discriminator.

\subsection{Corner plots in the output space}
\begin{figure*}[htpb]
    \centering
    \includegraphics[width=1\textwidth]{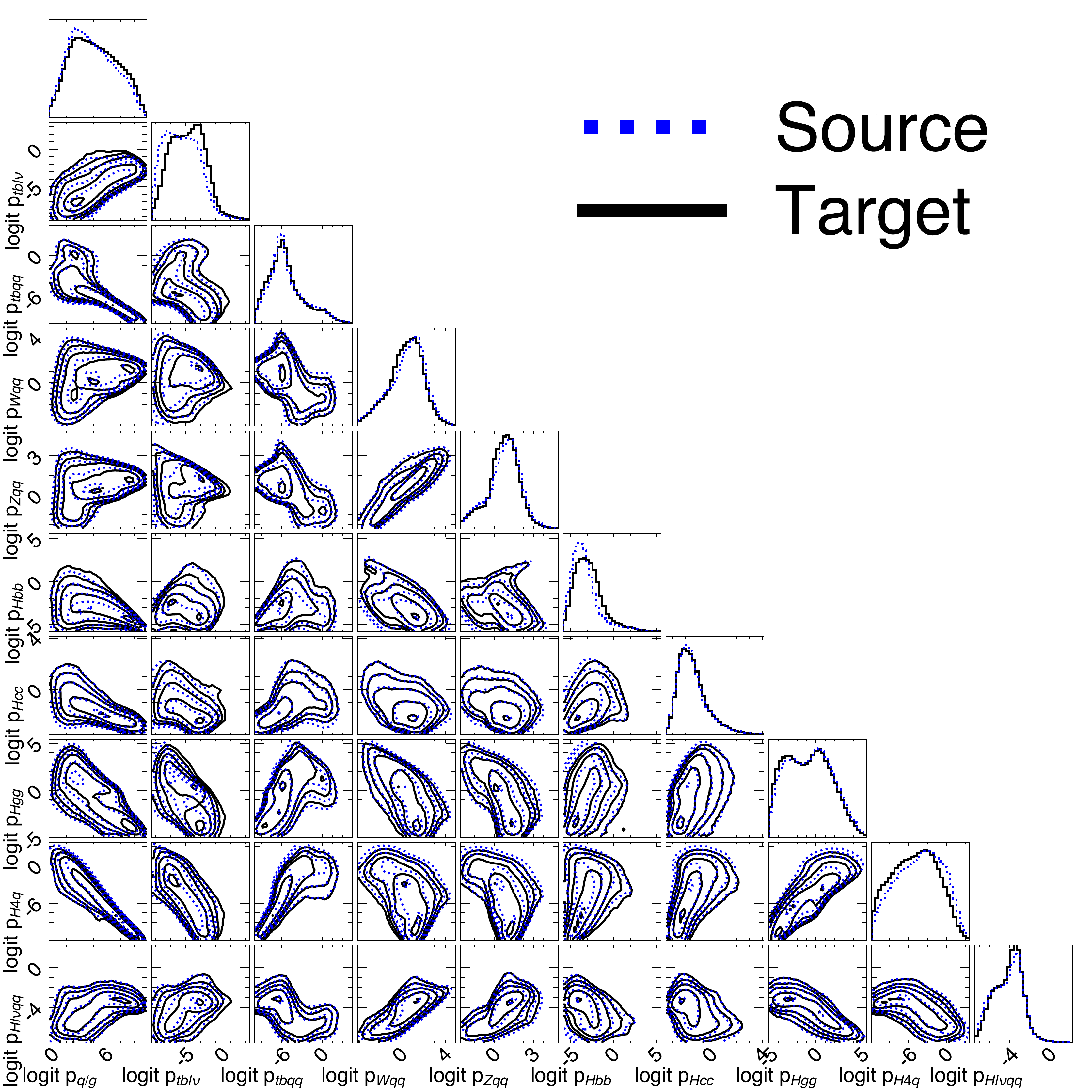}
    \caption{
    Comparisons of the marginal distribution of the output space $z_{10}^{'}$ of the classifier.
    The black distribution represents target distribution, red is the transported distribution and blue is the source distribution.
    The transport has been derived in $z_{128}$ and the later layers in the \OMLP have been applied to map it to $z_{10}^{'}$. 
    Closure of the calibration is indicated by the difference to the black distribution.
    Description of the labels can be found in \cref{eq:labels}.
    }
    \label{fig:pca_128_source}
\end{figure*}

\begin{figure*}[htpb]
    \centering
    \includegraphics[width=1\textwidth]{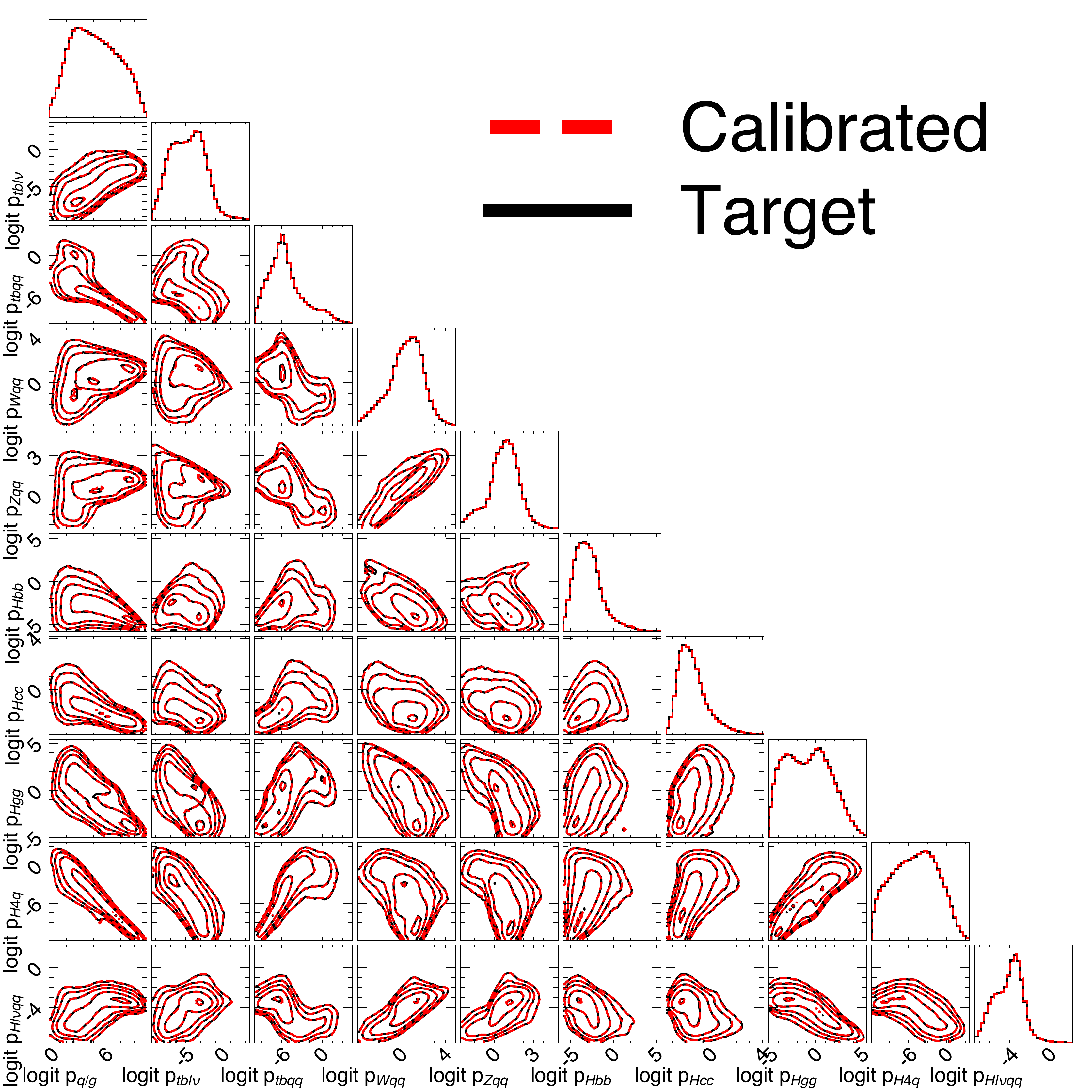}
    \caption{
    Comparisons of the marginal distribution of the output space $z_{10}^{'}$ of the classifier.
    The black distribution represents target distribution, red is the transported distribution and blue is the source distribution.
    The transport has been derived in $z_{128}$ and the later layers in the \OMLP have been applied to map it to $z_{10}^{'}$. 
    Closure of the calibration is indicated by the difference to the black distribution.
    Description of the labels can be found in \cref{eq:labels}.
    }
    \label{fig:pca_128_target}
\end{figure*}

\cref{fig:pca_128_source,fig:pca_128_target} show the corner plots of the output space of the classifier with and without calibration.

\clearpage

\section{Hyperparameters}\label{sec:hyperparameters}
\subsection{Classifier trained on the JetClass dataset}
Transformer based classifier trained on JetClass dataset. The output of the classifier is a 10-dimensional vector representing the probability of the jet belonging to one of the 10 classes.
The classifier was trained with 100.000 optimizer steps of AdamW~\cite{adamw} with a learning rate of $10^{-4}$ and a warmup of 10.000 steps. The full hyperparameters of the classifier are listed in \cref{tab:classifier_hyperparameters}.
\begin{table}[htpb]
    \centering
    \begin{tabular}{lll}
    \hline
    \hline
    \multirow{5}{*}{Train} & Number of epochs  & 4                   \\
                           & Gradient clipping & 1.0                 \\
                           & LR scheduler      & Warmup($10^{-4}$, 10000) \\
                           & Batch size        & 1000                \\ 
                           & Optimizer        & AdamW                \\ 
                           \hline
    \multirow{6}{*}{Transformer} & $d_{model}$               & 512             \\
                                 & Dropout                   & 0.1             \\
                                 & Number of encoders        & 4               \\
                                 & MLP                       & GLU             \\
                                 & Dimension increase in MLP & $\times$2              \\
                                 & Attention head            & 8               \\ \hline
    \multirow{2}{*}{\OMLP}   & Number of layers          & 5               \\
                                 & Activation function       & GELU            \\
                                 \hline
                                 \hline
    \end{tabular}
    \caption{Hyperparameters of the classifier trained on JetClass. 
    Diagram of the architecture can be seen in \cref{fig:transformer_architecture}.}
    \label{tab:classifier_hyperparameters}
\end{table}

\subsection{Hyperparameters of the ICNN}
Two ICNN networks are used to derive the optimal transport map,
both with the same hyperparameters, which are listed in \cref{tab:icnn_hyperparameters}.
The ICNNs are trained on source and target samples of size 8,181,000, with a validation
size of 909,000 for each source and target.
The latent space is standardized to have zero mean and unit variance before
being passed to the ICNNs.
The computational cost of training the ICNNs is around 48 hours on a mid-range
consumer GPU with 8 GB of memory. The ICNNs consist of restricted linear layers
and thus have a low memory footprint, but due to the minmax optimization, the
training process is long. Scaling to higher dimensional latent spaces would
require more memory and computational resources but is similar to scaling a
standard MLP. 
\begin{table}[htpb]
    \centering
    \begin{tabular}{lll}
    \hline
    \hline
    \multirow{1}{*}{nICNN} & Number of blocks    & 4                   \\ \hline
    \multirow{4}{*}{ICNN block} & Number of layers    & 2                   \\
                             & Hidden dimension    & 2048                 \\
                             & Activation function & Zeroed Softplus                \\
                             & Optimizer           & AdamW               \\ \hline
    \multirow{7}{*}{Train}   & LR Scheduler        & CosineAnnealingLR($5\cdot 10^{-4}, 10^{-7}$) \\
                             & Optimizer           & AdamW                 \\
                             & f per g            & 4                 \\
                             & g per f            & 10                 \\
                             & Steps per epoch     & 512                 \\
                             & Epochs              & 1000                 \\
                             & Batch size          & 1024                \\ 
                             \hline
                             \hline
    \end{tabular}
    \caption{Hyperparameters of the ICNNs used to derive the optimal transport map.}
    \label{tab:icnn_hyperparameters}
\end{table}

\subsection{Hyperparameters of the classifiers used for evaluation \DL}
In the evaluation stage of the latent space calibration, a classifier, \DL, is trained to distinguish between the target and source and calibrated and target. The hyperparameters of \DL are listed in \cref{tab:dl_hyperparameters}.
The total number of parameters in \DL is around a $1$ million.
\begin{table}[htpb]
    \centering
    \begin{tabular}{lll}
    \hline
    \hline
    \multirow{5}{*}{\DL network} & Number of layers    & 4                   \\
                             & Hidden dimension    & 512                 \\
                             & Activation function & SiLU                \\
                             & Normalization       & LayerNorm           \\
                             & Optimizer           & AdamW               \\ \hline
    \multirow{4}{*}{Train}   & LR Scheduler        & CosineAnnealingLR($10^{-4}, 10^{-7}$) \\
                             & Early stopping      & 100                 \\
                             & Epochs              & 200                 \\
                             & Batch size          & 1024                \\ 
                             \hline
                             \hline
    \end{tabular}
    \caption{Hyperparameters of the classifier \DL used to discriminate between source vs target and calibrated vs target distributions.}
    \label{tab:dl_hyperparameters}
\end{table}

\section{Input variables}\label{sec:jetfeat}

\cref{tab:jetclass_csts_features,tab:jetclass_features} list the features of the particles and jets in the JetClass dataset used to train the classifier.

\begin{table}[htpb]
    \centering
    \begin{minipage}{0.48\textwidth}
        \centering
        \begin{tabular}{ll}
        \hline
        \hline
        \multicolumn{2}{l}{Continuous features}          \\ \hline
        Transverse momentum             & $p_T$          \\
        Pseudorapidity to the jet axis  & $\Delta \eta$  \\
        Azimuthal angle to the jet axis & $\Delta \phi$  \\
        Transverse impact parameter     & $d_0$          \\
        Longitudinal impact parameter   & $z_0$          \\
        Uncertainty on $d_0$            & $\sigma_{d_0}$ \\
        Uncertainty on $z_0$            & $\sigma_{z_0}$ \\ \hline
        \multicolumn{2}{l}{Particle type index}          \\ \hline
        Photon                          & 0              \\
        Negative hadron                 & 1              \\
        Neutral Hadron                  & 2              \\
        Positive Hadron                 & 3              \\
        Electron                        & 4              \\
        Positron                        & 5              \\
        Muon                            & 6              \\
        Anti-muon                       & 7              \\ \hline\hline
        \end{tabular}
        \caption{Features of the particle constituents of the jets in the JetClass dataset. The preprocessing of the features followed the procedure described in Ref~\cite{MPM}.}
        \label{tab:jetclass_csts_features}
    \end{minipage}%
    \hfill
    \begin{minipage}{0.48\textwidth}
        \centering
        \begin{tabular}{ll}
        \hline
        \hline
        \multicolumn{2}{l}{Scalar features} \\ \hline
        Jet transverse momentum   & $p_T$   \\
        Jet pseudorapidity        & $\eta$  \\
        Jet azimuthal             & $\phi$  \\
        Jet multiplicity          & $N$     \\
        Jet energy                & $E$     \\ \hline\hline
        \end{tabular}
        \caption{Features associated with a jet in the JetClass dataset. The preprocessing of the features followed the procedure described in Ref~\cite{MPM}.}
        \label{tab:jetclass_features}
    \end{minipage}
\end{table}

\clearpage

    \phantomsection
    \addcontentsline{toc}{chapter}{References}
    \printbibliography[title=References]

\end{document}